\documentclass[traditabstract]{aa} 
\usepackage{epsfig}
\usepackage{graphicx}
\usepackage{xcolor}
\usepackage{txfonts}
\usepackage{amssymb}
\usepackage{natbib}       

\bibpunct{(}{)}{;}{a}{}{,}
\newcommand{\be}{\begin{equation}}
\newcommand{\ee}{\end{equation}}
\newcommand{\bea}{\begin{eqnarray}}
\newcommand{\eea}{\end{eqnarray}}

\newcommand{\kappae}{\kappa_{\rm e}}

\newcommand{\fc}{f_{\rm c}}
\newcommand{\xe}{x_{\rm e}}
\newcommand{\melectron}{m_{\rm e}} 
\newcommand{\mproton}{m_{\rm p}} 
\newcommand{\momentum}{\mbox{\boldmath $p$}}
\newcommand{\momcap}{\mbox{\boldmath $P$}}

\newcommand{\vecv}{\mbox{\boldmath $\varv$}}
\newcommand{\veloc}{\varv}

\begin{document}

\title{Accretion heated atmospheres of X-ray bursting neutron stars}
\author{
V.~F. Suleimanov\inst{1,2,3}
\and
J. Poutanen\inst{3,4,5}
\and 
K. Werner\inst{1}}


\institute{
Institut f\"ur Astronomie und Astrophysik, Kepler Center for Astro and
Particle Physics, Universit\"at T\"ubingen, Sand 1,  72076 T\"ubingen, Germany\\ \email{suleimanov@astro.uni-tuebingen.de}
\and Astronomy Department, Kazan (Volga region) Federal University, Kremlyovskaya str. 18, 420008 Kazan, Russia
\and Space Research Institute of the Russian Academy of Sciences, Profsoyuznaya str. 84/32, 117997 Moscow, Russia 
\and Tuorla Observatory, Department of Physics and Astronomy,  FI-20014 University of Turku, Finland 
\and Nordita, KTH Royal Institute of Technology and Stockholm University, Roslagstullsbacken 23, SE-10691 Stockholm, Sweden
}

\date{Received xxx / Accepted xxx}

   \authorrunning{Suleimanov et al.}
   \titlerunning{Accretion heated neutron star atmospheres}

\abstract
{Some thermonuclear (type I) X-ray bursts at the neutron star surfaces in low-mass X-ray binaries take place during hard persistent
 states of the systems. Spectral evolution of these bursts is well described by the atmosphere model of a passively cooling neutron
  star when the burst luminosity is high enough. The observed spectral evolution deviates from the model predictions when 
 the burst  luminosity drops below a critical value of 20--70\% of the maximum luminosity. 
The amplitude of the deviations  and the critical luminosity correlate with the persistent luminosity, which leads us to suggest that
 these deviations are induced by the additional heating of the accreted  particles. 
We present a method for computation of the neutron star atmosphere models heated by accreted particles assuming that their 
energy is released via Coulomb interactions with electrons.  We computed  the temperature structures and the emergent 
spectra of the atmospheres of various chemical compositions and investigate the dependence of the results on the velocity of  
accreted particles, their temperature and the penetration angle.  We show that the heated atmosphere develops two different 
regions. The upper one is the hot (20--100 keV) corona-like surface layer cooled by Compton scattering, and the deeper, almost 
isothermal optically thick region with a temperature of a few keV.  The emergent spectra correspondingly have two components:
 a blackbody with the temperature close to that of the isothermal region and a hard  Comptonized component (a power law with 
 an exponential decay). Their relative contribution depends on the ratio of the energy dissipation rate of the accreted particles 
 to the intrinsic flux from the neutron star  surface.  These spectra deviate strongly from those of undisturbed, passively cooling 
 neutron star  atmospheres, with the main differences being the presence of a high-energy tail and a strong excess in 
 the low-energy part of the  spectrum. They also lack the iron absorption edge, which is visible in the spectra of undisturbed 
  low-luminosity atmospheres with  solar chemical composition.  Using the computed spectra, we obtained the dependences
   of the dilution and color-correction factors as functions of relative luminosities for pure helium and solar abundance atmospheres. 
We show that the { helium model atmosphere heated by accretion corresponding to 5\%  of the Eddington luminosity}  
describes well the late  stages of the X-ray bursts in 4U\,1820$-$30.
}


\maketitle
%

\section{Introduction}

Type I thermonuclear X-ray bursts have been used to obtain neutron star (NS) parameters. 
In particular, bursts happening during the hard persistent states of the systems can be used for such analysis \citep[see discussions in][]{SPRW11,Kajava.etal:14,Poutanen.etal:14,Suleimanov.etal:16}.\footnote{We note here that the bursts occurring in the soft, high-accretion-rate state never show spectral evolution consistent with theoretical prediction  \citep{Kajava.etal:14}. Therefore, this kind of burst cannot be used to determine NS parameters at all. } 
In order to use full information on the variations of the burst spectrum temperature and normalization, the so called cooling tail method or its modification \citep{SPRW11,Poutanen.etal:14,Suleimanov.etal:17} have been used. 
More accurate results can be obtained by directly fitting the burst spectra at different flux levels with NS atmosphere models \citep{Nattila.etal:17}.
However, these hard-state  bursts show a spectral evolution that deviates from the theoretically predicted  behaviour for passively cooling NSs when the burst luminosity drops below a certain level.
Furthermore, the higher the persistent flux, the more significant are the deviations. 
For instance, deviations start to be visible at burst luminosities below 20\% of the Eddington value $L_{\rm Edd}$ for the bursts taking place at persistent luminosities of about $0.01\,L_{\rm Edd}$ \citep{Nattila.etal:16}. 
On the other hand, the X-ray bursts of the helium accreting NS in 4U\,1820$-$30 happen at the relatively high persistent luminosity of about $0.07\,L_{\rm Edd}$, and such deviations begin at significantly higher burst luminosity of $\approx 0.7\,L_{\rm Edd}$ \citep{Sul.etal:17}. 
This leads us to the conclusion that additional heating by the accreted gas is the cause of the deviations. 

Therefore we need to develop NS atmosphere  models with an additional energy dissipation in the surface layers. 
Such  models would be applicable to the hard-state bursts. 
On the other hand, in the soft state the accreted matter likely  spreads over the NS surface { forming} a spreading/boundary layer \citep{IS99,SulP06}. 
Its interaction with the NS atmosphere cannot be described by a formalism involving interaction of single particles, but a hydrodynamical treatment would be needed.  
Thus the modelling of bursts happening in the soft state will not be considered here. 

X-ray spectra of accreting compact objects such as NSs and black holes in their hard persistent states form in a hot rarefied medium.  
In black holes, the most likely source of this radiation is the inner  geometrically thick and optically thin  hot accretion flow \citep[see e.g. reviews by][]{DGK07,PV14,YN14}. 
In accreting NSs  this radiation maybe produced also at the NS surface \citep{DDS01}, which even can dominate the total power. 
In this paper we consider old weakly magnetized NSs in  low-mass X-ray binaries (LMXBs).

\citet{ZS69} were among the first to discuss the emission from a NS surface heated by accreted matter. 
They considered two physical processes that lead to deceleration of protons  in NS atmospheres. 
The simplest one is  Coulomb interaction with electrons.
In this case, the main part of the proton kinetic energy is released deep in the optically thick atmospheric layers resulting in the rather thermal emergent spectra. 
Another possibility is excitation of collective plasma processes by the protons. 
Here the proton energy is dissipated within an optically thin atmospheric layer and the emergent spectrum may have a shape far { different} from the blackbody. 
\citet{AW73} confirmed these conclusions using first numerical models of accretion heated NS atmospheres. 
They considered hydrogen NS atmospheres heated by protons falling radially with free-fall velocity.
The proton deceleration was considered in a self-consistent way. 
Two input parameters of the model were considered, namely the NS mass and the accretion luminosity. 
The most important properties of the accretion heated atmospheres, such as an overheated Compton-cooled outer layers and spectral 
hardening with the increase of the accretion luminosity were found in this work.

Later  the problem was considered in detail by \citet{BSW92}. 
They provided useful relations for describing the proton velocity profiles and the vertical distribution of the accretion heating, which 
were used by other authors for modeling accretion heated atmospheres  \citep{Turolla.etal:94, Zampieri.etal:95, Zane.etal:98}. 
They also considered hydrogen atmospheres heated by accreting protons. 
They confirmed the basic properties of the heated model atmospheres found by \citet{AW73}, and additionally declared existing 
of the so called hot solutions, where the temperature can reach almost 10$^{12}$\,K \citep{Turolla.etal:94}. 
The properties of such "hot" solutions were investigated by  \citet{Zane.etal:98}, who, in particular, demonstrated the importance of electron-positron pair creation. 
\citet{Zampieri.etal:95} concentrated on the low-luminosity accretion heated atmospheres. 
They assumed the energy release mainly in the optically thick layers of the atmosphere. 
Therefore, the temperature structures of these models at the  spectrum formation depths were close to that of the undisturbed model atmospheres with the same luminosities. 
As a result, they obtained spectra  harder than the blackbody and very similar to those of   undisturbed NS  atmospheres \citep[see, e.g.][]{Romani:87, Zavlin.etal:96}. 
The main reason is the dependence of the free-free opacity on the photon energy $k_{\rm ff} \sim E^{-3}$. 
Photons prefer to escape from an atmosphere in the most transparent energy bands. 
\citet{Zampieri.etal:95}  also found a clear division between the outer heated atmospheric layers cooled by Compton scattering from the  almost isothermal inner part cooled by free-free processes.

\citet{DDS01} computed the accretion heated NS atmospheres and proton deceleration self-consistently and used the results for interpretation of the hard persistent spectra of some LMXBs. 
For the first time they showed the importance of the temperature of accreted protons on  the atmosphere properties. 
They also considered  the bulk velocity as a free parameter. 

The results obtained by \citet{Zampieri.etal:95} and \citet{DDS01} were widely used for interpretation of the NS spectra in LMXBs during low mass-accretion-rate states  \citep[see e.g.][]{Homan.etal:14, Wijnands.etal:15}. 
Also deceleration of the protons in magnetized NS atmospheres was considered before  \citep[see e.g.][]{NSW93}, and the results were used for the modeling of such atmospheres by \citet{Zane.etal:00}.

In this paper we develop a method to compute models of NS atmosphere heated by accreted matter.  
It is based on the approach developed by \citet{DDS01}, which we slightly modified.
In contrast to previous papers devoted to pure hydrogen atmospheres heated by protons, here we consider arbitrary chemical 
composition of the atmospheres and an arbitrary mix of protons and $\alpha$-particles for the accreted particles. 
We present the basic properties of the accretion-heated NS atmospheres and their dependence on the input parameters such as 
the velocity of accreted particles  and their direction, the accretion- and the intrinsic NS luminosities. 
We also compare the developed models to the observed spectral evolution of hard-state X-ray bursts.
 
\section{The model}
\label{sec:method}
\subsection{Main equations}

We consider steady-state hot NS atmospheres in plane-parallel approximation with additional heating in the surface layers caused by supra-thermal accreted particles.
The basic  principles of the modeling of hot, undisturbed NS atmospheres were presented in our earlier works  \citep{SPW11,SPW12}.
This work is  based on the formalism described in the latter paper.

We consider a uniform non-rotating NS with mass $M$, radius $R$ and intrinsic luminosity $L$.
The input parameters for the NS atmosphere model are the surface gravity $g$,
the effective temperature $T_{\rm eff}$, and the chemical composition.
The effective temperature here is a parameter describing the intrinsic bolometric surface flux 
\be
   \sigma_{\rm SB}T_{\rm eff}^4 =  \int_0^\infty F_{0,\nu} (0) \, d\nu = F_0 ,
\ee
where $ \sigma_{\rm SB}$ is the Stefan-Boltzmann constant. 
Here the argument of $F_{0,\nu}(0)$ means that the flux is measured at the surface, while the lower index $0$ 
corresponds to the intrinsic flux.  
 Another parameter, the relative intrinsic NS luminosity $\ell = L/L_{\rm Edd}$, could be used instead of $T_{\rm eff}$.
Here and below a relative luminosity means some luminosity normalized to the Eddington luminosity. 
Both the surface gravity $g$ and the Eddington luminosity $L_{\rm Edd}$ are computed for Schwarzschild metric  \citep[see details in][]{Suleimanov.etal:16, DS18}. 
The Eddington luminosity depends also on the hydrogen mass fraction $X$.

In the previous works \citep{SPW11, SPW12,Nattila.etal:15}, we computed  models of atmospheres  consisting of pure hydrogen, pure helium,  solar hydrogen/helium mixture enriched by various fractions of heavy elements, from the solar abundance  down to an one-hundredth of that,  as well as heavy metals. 
Here we consider pure helium and solar composition NS atmospheres only. 
We express the additional luminosity arising due to external heating by fast particles through the relative accretion luminosity  $\ell_{\rm a} = L_{\rm a}/L_{\rm Edd}$. 
We assume that the kinetic energy of the fast particles is thermalized  inside the atmosphere and is radiated  away through the surface. 
As a result, the emergent bolometric flux is higher by this relative fraction: 
\be
F=    \int_0^\infty F_\nu(0)\,d\nu = F_0 (1+\ell_{\rm a}/\ell).
\ee 

Let us consider  for simplicity identical particles with  mass $m_{\rm i}$ moving along the normal to the NS surface with the same velocity $\veloc_0$ before penetrating into the atmosphere (more general cases  will be described later).  
The total accretion luminosity is 
\be   \label{accrL}
       L_{\rm a} =   4\pi R^2 \,\dot m c^2\,(\gamma-1),  
\ee
where 
\be
      \gamma=\frac{1}{\sqrt{1-\veloc_0^2/c^2}} 
\ee
is the Lorentz factor. 
Here $ \dot m$ is the  local mass accretion rate (per unit area), which can be related to $\ell_{\rm a}$ and the NS surface gravity $g$  as 
\be \label{dotm}
\dot m = \ell_{\rm a}\frac{g}{\kappae\,c\,(\gamma-1)}. 
\ee  
       
The value of $\dot m$ is conserved in the atmosphere, but the number density $n$ of the accreted particles and their velocity 
$\veloc$ can change according to the mass conservation law
\be
   \dot m = n m_{\rm i} \veloc.
\ee
The flux of the accreting particles creates the ram pressure $P_{\rm ram} = \dot m\, \gamma \veloc$, which manifests itself as an additional acceleration pressing the atmosphere down  
\be \label{eq:gram}
       g_{\rm ram} = \frac{1}{\rho}\frac{dP_{\rm ram}}{dr} = -\dot m \frac{d}{dm}(\gamma \veloc), 
\ee
where $\rho$ is the plasma density of the atmosphere and $m$ is a Lagrangian independent variable--column density, defined as $dm=-\rho dr$. 
We note that the velocity $\veloc$ is positive in the direction of increasing $m$.
The ram pressure force has to be included into the hydrostatic equilibrium equation, the first equation 
determining  the structure of the atmosphere 
\be
  \frac{1}{\rho}\frac{dP}{dr} = -g-g_{\rm ram}+g_{\rm rad},
\ee
where $P$ is the gas pressure and $g_{\rm rad}$ is the radiative acceleration (as defined in \citealt{SPW12} for non-isotropic Compton scattering).

The accreted particles do not directly affect the equation of radiative transfer  for the specific intensity $I(x,\mu)$, and it is the same as for the undisturbed atmosphere  \citep[see details in][]{SPW12}.
 Here $\mu = \cos \theta$ is the cosine of the angle between the surface normal  and the direction  of radiation propagation, 
$x=h\nu/\melectron c^2$ is the photon energy in units of electron rest mass. 
We note that the fully relativistic, angular dependent redistribution function is used to describe  Compton scattering
 \citep[see][]{NP93, PS96}. 

The accretion of fast particles heats the NS surface producing additional radiation flux  $\dot m c^2 (\gamma-1)$. 
The kinetic energy of fast particles is released in the NS atmosphere gradually and the local energy dissipation rate 
has to be proportional to the deceleration rate of the particles considered in the next section. 
The general form of the local energy dissipation rate, which is important for modeling of the heated atmospheres, takes the following form:
\be \label{eq:Qplus}
 Q^+ = -\frac{d}{dm} \left(\dot m c^2 (\gamma-1)\right) =- \dot m \gamma^3 \veloc  \frac{d\veloc}{dm}.
\ee
We assume that all the local dissipated energy is transformed to radiation, therefore, we can present the change in the radiation flux  as
\be
Q^+  = -\frac{d}{dm} \int_0^\infty F_x(m)\,dx .
\ee
The energy balance equation now takes the form  
\be  \label{eq:econs}
2\pi\!  \int^{\infty}_{0}\!\!\! \!  dx\,  \!\int^{+1}_{-1} \!\! \!\left[\sigma(x,\mu) + k(x)\right]  \left[I(x,\mu)-S(x,\mu)\right] \, d\mu = -Q^+,
\ee
and has to be fulfilled at the every depth in the atmosphere (for the undisturbed atmosphere $Q^+=0$). 
Here  $k(x)$ is  the ``true'' absorption opacity,  and  
$\sigma(x,\mu)$ is the electron scattering opacity \citep[see definitions in][]{SPW12}.

Thus the hydrostatic equilibrium  and the energy balance equations  describing  the NS atmosphere,  can be modified easily 
to account for accretion of fast particles with equal mass and velocity along the normal. 
It is clear that the dependence { of the particle deceleration $d\veloc/dm$ on depth} fully determines the atmosphere model (in addition to the standard parameters defining the undisturbed atmosphere). 
Below we derive the deceleration function and generalize the treatment to a distribution of particles obliquely accreting to the NS surface.
 
\subsection{Stopping of fast particles  in the atmosphere}

Let us consider a fast particle having velocity $\veloc$ directed into the atmosphere, mass $m_{\rm i}$ and  charge $Ze$ moving through the plasma with temperature $T$  and electron number density $n_{\rm e}$. { Here and further $e$ is the charge of electron and $\melectron$ is the mass of electron.}
The particle loses energy via Coulomb interactions with the plasma electrons. 
The basic physics  of deceleration of such particles is well established (see e.g. \citealt{AW73, DDS01}, and chapter 3 in \citealt{FKR02}).
The main parameter describing deceleration rate is the slowing-down timescale
\be \label{tsd}
      t_{\rm s} = -\frac{\veloc}{d\veloc/dt} \approx 
      \frac{m_{\rm i}^2 \veloc^3}{4\pi n_{\rm e}(Ze)^2e^2\,\ln \Lambda}
      \frac{\melectron}{m_{\rm i}+\melectron},
\ee 
which { can be solved for} the deceleration  
\be \label{dfc}
\frac{d\veloc}{dt}= -\frac{\veloc}{t_{\rm s}} \approx -f(\xe)\,\frac{4\pi n_{\rm e}(Ze)^2e^2}{m_{\rm i} \melectron \veloc^2}\,\ln \Lambda.
 \ee
The corresponding energy loss is
\be \label{eq:dedt}
       \frac{dE}{dt} \approx -f(\xe)\,\frac{4\pi n_{\rm e}Z^2 e^4}{\melectron \veloc}\,\ln \Lambda.
\ee
Here  $\ln \Lambda$ is the Coulomb logarithm:
\be \label{qlog}
       \ln \Lambda \approx \ln \frac{2m_{\rm i} \melectron^{3/2} \veloc^2}{\hbar (m_{\rm i}+\melectron)\sqrt{4\pi n_{\rm e} e^2}}
\ee
for the case $\veloc > \alpha\,c \approx c/137$ \citep{Larkin60} ($\alpha$ is the fine-structure constant). 
At low particle velocities ($\veloc \ll \alpha\,c$), Eq.\,(\ref{qlog}) gives too small values and we use instead the standard Coulomb logarithm for a thermal plasma 
\be
       \ln \Lambda_{\rm T} \approx  \ln \frac{3}{2e^3}\frac{(kT)^{3/2}}{(\pi\,n_{\rm e})^{1/2}},
\ee  
 where $k$ is Stefan-Boltzmann constant.
In computations we choose the { larger value of} $\ln \Lambda$ and  $\ln \Lambda_{\rm T}$. 
 
The function $f(\xe)$ takes into account the reduction of the deceleration force  at low particle velocities and is expressed as
\be
 f(\xe) \approx \phi(\xe)-\xe \phi'(\xe),
\ee
where $\phi(x)$ is the error function, and $x_{\rm e} = (\melectron \veloc^2/kT)^{1/2}$ is the ratio of the particle velocity to the averaged thermal electron velocity.  
For the computations we use the approximation 
\be
     f(x_{\rm e}) \approx \frac{4\xe^3}{3\sqrt{\pi}+4\xe^3}
\ee
suggested by \citet{AW73}.

\subsection{Heating by { thermally} distributed fast particles}

The X-ray spectra of LMXBs in their hard persistent states are well described with the spectrum of radiation Comptonized in a hot electron 
slab with optical depth $\tau_{\rm e} \sim 1$ and temperature $kT_{\rm e} \sim 15-50$\,keV \citep{Barret00,BGS17}. 
This radiation may be produced in the optically thin and geometrically thick accretion flow which could be described
 with the advection-dominated accretion flow (ADAF) model \citep[see e.g.][]{YN14}. 
The ADAF model differs from the standard accretion disc model \citep{SS73} at least in two respects. 
The gravity of the central object is balanced not only by the rotation, but the radial gas pressure gradient as well. 
It is also important that the ion temperature in ADAF model is close to the virial one at a given radius $r$
\be
     T_{\rm vir}(r) \approx \frac{GMm_{\rm i}}{3k\,r} \approx 5.4\times 10^{10}\,{\rm K}\,\left(\frac{M}{M_\odot}\right)\,
     \left(\frac{r}{10^7\,{\rm cm}}\right)^{-1}\,\left(\frac{m_{\rm i}}{\mproton}\right)
\ee
and can reach $10^{12}$\,K in the vicinity of the NS. 
Such high ion temperature provides a significant accretion flow thickness $h$, comparable with the radial distance,
 $h/r \sim 1$, and a high radial velocity of the matter,  comparable with the free-fall velocity $V_{\rm ff}(r) = c\sqrt{R_{\rm S}/r}$. 
These properties imply a quite large impact angle of the { accreting} ions to the NS surface normal, and we consider  this angle
 $\Psi$ as a parameter. Another parameter is the ratio of the bulk velocity of the particles to the free-fall velocity at the NS surface $\eta= V_0 / V_{\rm ff}(R)$ \citep[we follow here][]{DDS01}.  
Because the mean ion velocity at the virial temperature is close to $V_{\rm ff}$, the distribution of the ions over velocities has to be taken into account \citep{DDS01}. 
The ADAF could be so hot and rarefied that the Coulomb interactions are not efficient enough to establish the Maxwellian distribution \citep{MQ97}. 
Thus potentially the ion velocity distribution  may significantly deviate from the Maxwellian. 
In this work, however,  we assume that ions follow the relativistic Maxwellian distribution  and we describe the ion temperature through its ratio to the virial temperature, $\chi=T_{\rm i}/T_{\rm vir}(R)$. 
  
It is necessary to note that in addition to the bulk kinetic energy, thermally distributed accreted particles contribute their thermal energy (enthalpy) to the atmosphere. 
Therefore, the accretion luminosity is larger than that given by Eq.\,(\ref{accrL}): 
\be \label{therm_contr}
      L_{\rm a} \approx 4\pi R^2\, \dot m \,\left(c^2(\Gamma-1)+\frac{5}{2}\frac{kT_{\rm i}}{m_{\rm i}}\right),
      \ee
      where
      \be
      \Gamma=1/{\sqrt{1-V_0^2/c^2}} 
\ee
is the bulk Lorentz factor. 
The contribution of thermal energy is important at $\chi > 0.1$.  
We note that the input parameter in our model is a relative accretion luminosity $\ell_{\rm a}$, and therefore  in our numerical computations we correct the local mass accretion rate obtained using Eq.\,(\ref{dotm}) to keep $\ell_{\rm a}$ fixed.
The second important thing is that the parameters $\eta$ and $\chi$ are not completely free. 
The total gravitational potential energy of the accreted matter related to the free-fall Lorentz factor $\Gamma_{\rm ff}$ as $\dot m c^2(\Gamma_{\rm ff}-1)$  is released by three means. 
It is partially radiated away by the hot accretion flow, with the corresponding flux $F_{\rm flow}$. 
The rest is split up between the enthalpy of the hot matter and the energy of the bulk motion. 
In non-relativistic approximation, { the latter} can be presented by the equation $\dot m V^2_{\rm ff}/2 \approx F_{\rm flow}+\eta^2\dot m V^2_{\rm ff}/2 +\chi \dot m V^2_{\rm ff}/2$.   
Thus we have an additional constraint on the parameters $\eta^2+\chi < 1$.   

Furthermore, for the isotropic particle distribution in the co-moving frame  there is a significant number of particles penetrating into the atmosphere and contributing their energy even when the bulk velocity is parallel to the surface, when the contribution of the bulk kinetic energy formally is absent.  
This effect is especially significant for the ram pressure. 
We account for this effect in our calculations.

The relativistic generalization of the Maxwellian distribution is the Maxwell-J\"uttner distribution in the co-moving frame 
\be \label{MJd}
        f'(p') =\frac{\Theta}{4\pi\,K_2(\Theta)}\,\exp{\left(-\Theta\gamma' \right)},
\ee
where $\gamma'$ is the particle Lorentz factor 
\be
       \gamma'=\frac{1}{\sqrt{1-\veloc'^2/c^2}} = \sqrt{1+\momentum'^2},
       \ee
$\veloc'$ is the particle velocity, $\momentum'= \gamma' \vecv'/c$ is the dimensionless  momentum, $p'=|\momentum'|$, $\Theta$ is the normalized Gibbs parameter  $\Theta = m_{\rm a}c^2/kT_{\rm i}$, and $K_2(\Theta)$ is the Macdonald function (the modified Bessel function of the second kind).
The distribution is normalized to unity: 
\be \label{norm}
       \iiint\limits_{-\infty}^{+\infty}
\,f'(p'_{x},p'_{y},p'_{z})
       \,dp'_{x}\,dp'_{y}\,dp'_{z} = 
         4\pi\int\limits_{0}^{\infty}p'^2f'(p')\,dp' = 1 .      
\ee
When the gas moves with some bulk velocity defined by the dimensionless  average momentum $\momcap$ and corresponding Lorentz factor $\Gamma=\sqrt{1+\momcap^2}=(1-V_0^2/c^2)^{-1/2}$,  the Maxwell-J\"uttner distribution takes the form \citep{BB56,NP94a}
\be \label{fdsh}
         f(\momentum) = \frac{\Theta}{4\pi\,K_2(\Theta)\Gamma} 
       \exp{\left[-\Theta\left(\gamma\Gamma -\momentum \cdot \momcap\right)\right]}.
\ee

\begin{figure}
\centering
\includegraphics[angle=0,scale=1.0]{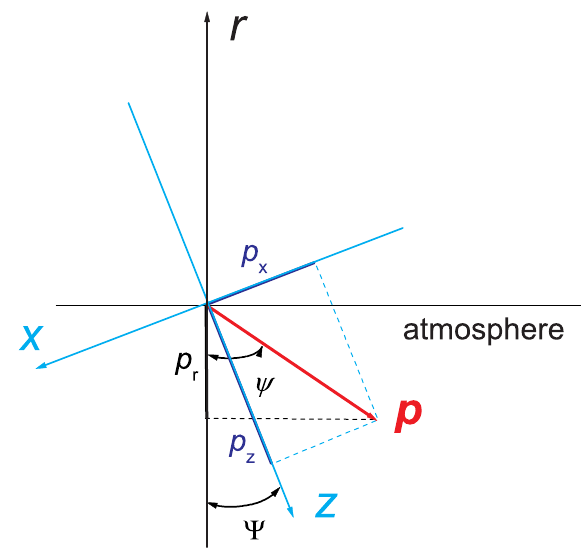}
\caption{\label{fig1}
The coordinate systems used in the paper. 
We also show the momentum  of a single particle and its components with the corresponding angles.
}
\end{figure}

Let us introduce two Cartesian coordinate systems with $z$ axis directed along the bulk velocity of the fast particles (see Fig.\,\ref{fig1}). 
The first one moves with the bulk velocity (the co-moving frame), and the second one (laboratory) is connected with the atmosphere. 
In the co-moving frame, the particle distribution follows the Maxwellian-J\"uttner distribution  (\ref{MJd}). 
In the laboratory frame, two components of the momentum, $p_{x}=p'_{x}$ and $p_{y}=p'_{y}$, 
retain their values, but the third one is transformed as
\be \label{pz}
     p_{\rm z} = \Gamma \left(p'_{z}+\frac{V_0}{c}\gamma'\right).
\ee
We note that the normalization of the distribution (\ref{fdsh}) remains unchanged 
\be
  \iiint\limits_{-\infty}^{+\infty}\,f(p_{x},p_{y},p_{z})
       \,dp_{x}\,dp_{y}\,dp_{z} = 1 . 
\ee
We note that the Lorentz factor $\Gamma$ in the denominator of (\ref{fdsh}) appears due to the Lorentz contraction of the elementary volume ($dz=dz'/\Gamma$). 
  
To estimate the energy deposition rate, we consider the impact of the particles distributed over the momentum components according to Eq.~(\ref{fdsh}), and assume that they move independently inside the atmosphere. 
Let us consider a particle with the momentum $\momentum=(p_{x},p_{y},p_{z})$. 
It has the {initial} velocity
\be
        \veloc_0=c\frac{p}{\sqrt{1+p^2}}
\ee
and moves along the line that makes angle $\psi$ with the surface normal (directed toward the surface):  
\be
       \cos\psi = -\frac{p_{r}}{p},
\ee
where $p_{r} = -(p_{z}\cos\Psi+p_{x}\sin\Psi)$, see Fig.\,\ref{fig1}. 
For the steady state flux of accreting particles we can replace the total time derivative in Eq.~(\ref{eq:dedt}) with the space derivative along the displacement direction $s$, and compute the deceleration 
\be
       \frac{dE}{dt} = \veloc\frac{dE}{ds}=m_{\rm i}\veloc \cos\psi\frac{d}{dr}\left(\gamma c^2\right)
       =-\rho \,\gamma^3\cos\psi\,m_{\rm i}\veloc^2 \frac{d\veloc}{dm}.
\ee
Here we used a relativistic representation for the kinetic energy of the particle  $\gamma m_{\rm i} c^2$ and replaced the derivative 
from Euler ($r$) to Lagrangian ($m$) coordinates.
Finally, using the energy dissipation rate due to Coulomb interaction given by Eq.~(\ref{eq:dedt}) we get
\be \label{dvdz}
    \frac{d\veloc}{dm} \approx -f(x_{\rm e})\frac{4\pi n_{\rm e}Z^2e^4}{\rho\,\gamma^3\cos\psi\,m_{\rm i} \melectron \veloc^3}\,
    \ln \Lambda.
\ee
For a given  atmosphere model (with given $T(m)$, $\rho(m)$ and $n_{\rm e}(m)$), we solve this equation to find $\veloc(m)$. 
The solution is then used to determine the  specific energy deposition rate per unit mass (see also Eq.\,\ref{eq:Qplus}):
\be
          q^+(p_{x},p_{y},p_{z})= -\gamma^3 \veloc \frac{d\veloc}{dm}  \approx f(x_{\rm e})\frac{4\pi n_{\rm e}Z^2 e^4}
          {\rho \cos\psi\,m_{\rm i} \melectron \veloc^2}\,\ln \Lambda.
\ee
Integrating this equation over the possible initial momenta (with $p_{r}<0$), we get the total energy deposition rate by accretion: 
\be \label{eq:Qplusfin}
      Q^+_{\rm a}= \dot m \iiint\limits_{-\infty}^{+\infty}\, \,f(p_{x},p_{y},p_{z})\,q^+ dp_{x}\,dp_{y}\,dp_{z}.   
\ee
The total energy released in the atmosphere is found by the integration over the depth 
\be
        F_{\rm a} = \int_0^\infty Q^+_{\rm a}(m)\,dm. 
\ee

Similarly, we compute the contribution per unit mass to the ram pressure force along
 $r$ (see Eq.\,\ref{eq:gram}):
\be
\cos\psi \frac{d}{dm}(\gamma\veloc) =- \cos\psi\, \gamma^3 \frac{d\veloc}{dm}\approx  
f(x_{\rm e})\frac{4\pi n_{\rm e}Z^2e^4}{\rho\,m_{\rm i} \melectron \veloc^3}\, 
    \ln \Lambda.
\ee
We note that only the normal component of the force is important.
Thus the total ram pressure acceleration is obtained by integrating over all momenta (with $p_r<0$)
\be \label{gram}
      g_{\rm ram} = - \dot m \iiint\limits_{-\infty}^{+\infty}     \,f(p_{x},p_{y},p_{z})\,\cos\psi\, \gamma^3 \frac{d\veloc}{dm}  
      \,dp_{x}\,dp_{y} \,dp_{z}.
\ee
 Here and before we determine the contributions to the energy dissipation rate and ram pressure of fast particles per unit mass 
 in order to separate the mass accretion rate $\dot m$ in the final equations (\ref{eq:Qplusfin}) and (\ref{gram}).

\subsection{Thermal conduction}

Previous works \citep{AW73,DDS01} demonstrated that the temperature gradient in the accretion-heated atmospheres can be significant, and the energy transport via thermal conduction may be not negligible. 
The thermal conduction flux is
\be\label{eq:fluxc}
         F_{\rm c}= \rho \kappa_{\rm c} \frac{dT}{dm},
\ee
where 
\be \label{eq:kappac}
   \kappa_{\rm c} = 1.85\times 10^{-5}\frac{T^{5/2}}{\ln \Lambda_{\rm T}} 
\ee   
is the heat transfer coefficient. 
The heating/cooling rate of the matter due to thermal conduction is given by
\be \label{Qc}
       Q_{\rm c} = \frac{dF_{\rm c}}{dm}.
\ee
The total heating rate in Eq.\,(\ref{eq:Qplus})  then becomes  
\be \label{allQ}
       Q^+ = Q^+_{\rm a} + Q_{\rm c}. 
\ee

\subsection{Computation of atmosphere structure}

The general method of computation of the atmosphere structure is the iterative temperature correction approach. 
It was described in our previous paper \citep[see Sect.~2 of][]{SPW12}. 
Here we report only the differences caused by  the inclusion of additional energy dissipation in the atmosphere. 
As in \citet{SPW12}, we start iterations with the gray atmosphere model  defined on a grid of Rosseland optical depths. 
Other  physical quantities such as gas pressure,  density,  number densities of the  ions and atoms included into the model, 
opacities, and the column density grid $m$ are computed for the given temperature distribution on the Rosseland optical depth
 grid. 
Further computations are performed on the column density grid $m$.
Using the current model atmosphere we compute the accretion and the conductivity heating using the method described above.
We also derive the { vertical} radiation flux distribution { through} the atmosphere. 
As we have not accounted for the contribution of the thermal energy of the accretion flow (see Eq.\,\ref{therm_contr}) 
when we evaluated the local mass accretion rate (Eq.\,\ref{dotm}), the obtained additional emergent flux $F_{\rm a}$ 
differs from the value determined by the input parameter  $\ell_{\rm a}$, namely $\ell_{\rm a}F_0/\ell$. 
Therefore, we correct the mass accretion rate using the correction factor $C$ determined by 
\be \label{corr_fct}
   \dot m_{\rm c} = C \,\dot m = \frac{\ell_{\rm a}F _0}{\ell F_{\rm a}}\,\dot m.
\ee
 The ram pressure acceleration (\ref{gram}) as well as the energy generation rate (\ref{eq:Qplusfin}) are corrected using 
$\dot m_{\rm c}$ instead of $\dot m$.

Then we solve the radiation transfer equation with Compton scattering taken into account \citep[see details in][]{SPW12} for the current model atmosphere. 
The equation is solved on the chosen photon energy grid covering all the range where the atmosphere radiates.  
In a self-consistent atmosphere  the integral radiation flux at every depth 
 \be
    F(m)=\int\limits_0^\infty F_x(m)\,dx
 \ee
has to be equal to the values determined before known from the physics of the atmosphere.  
For instance, the integral flux has to be constant  over the depth in an undisturbed atmosphere. 
In our case the integral total flux (a sum of the radiation flux and the flux due to thermal conductivity ($F(m)+F_{\rm c}(m)$) at every depth is determined by two terms: the intrinsic radiation flux and  the flux generated  by accretion at depths deeper than the considered depth. 
Therefore, the final radiation flux distribution over the depth is determined by the following equation
\be \label{int_energy}
       F(m) = F_0+ C\left(F_{\rm a} - \int\limits_0^m Q^+_{\rm a} (m')\,dm' \right) - F_{\rm c}(m).
\ee
Here we also take into account the correction to the mass accretion rate.
In fact, this flux distribution is the integral form of the energy balance equation.  
It has to be fulfilled together with the differential form of the energy balance equation (\ref{eq:econs}).

\begin{figure}
\centering
\includegraphics[angle=0,scale=1.0]{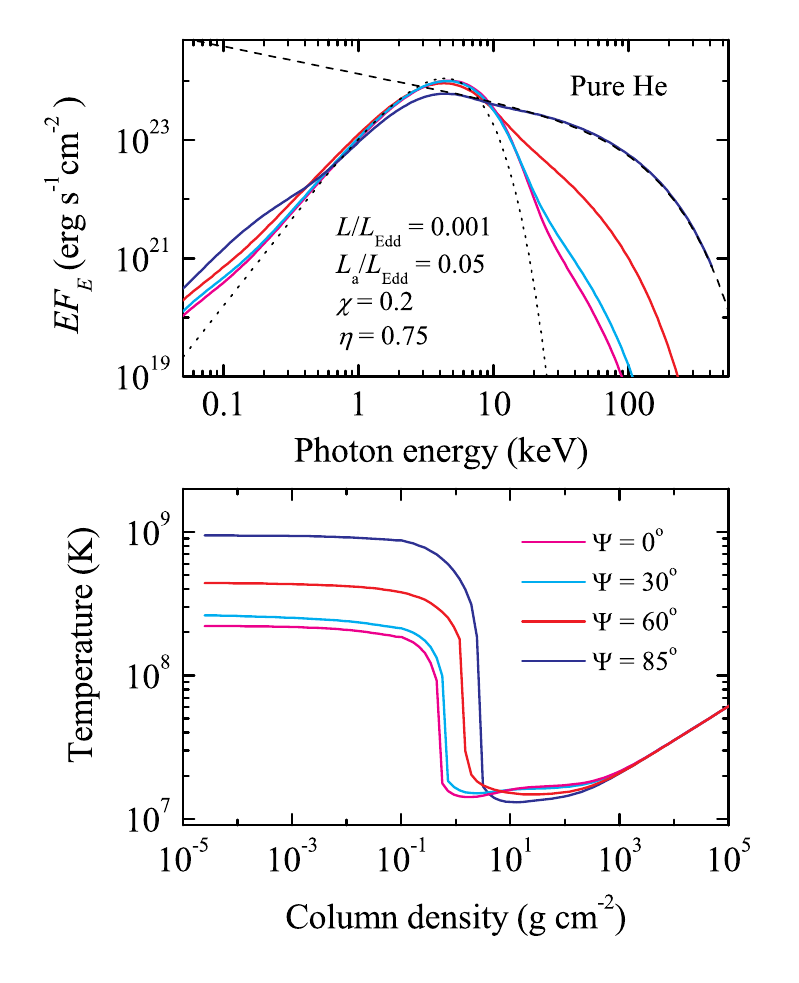}
\caption{\label{fig2}
The emergent energy spectra (top panel) and the temperature structures (bottom panel) for four low-luminosity ($\ell=0.001$)
helium model atmospheres heated by $\alpha$-particles with the  accretion luminosity $\ell_{\rm a}=0.05$ and
parameters $\eta$=0.75 and $\chi=0.2$. 
The models correspond to different incoming angles $\Psi=0\degr, 30\degr, 60\degr$, and 85\degr.
The emergent spectrum of the most inclined ($\Psi=85\degr$) particle flow is well fitted  at photon energies higher than 5 keV with an exponential cutoff power-law model with photon index $\Gamma=2.45$ and cutoff energy $E_{\rm cut}=85$ keV (dashed curve).  
The blackbody spectrum of temperature $kT=1.1$ keV is also shown by the dotted curve. 
}
\end{figure}

\begin{figure}
\centering
\includegraphics[angle=0,scale=1.0]{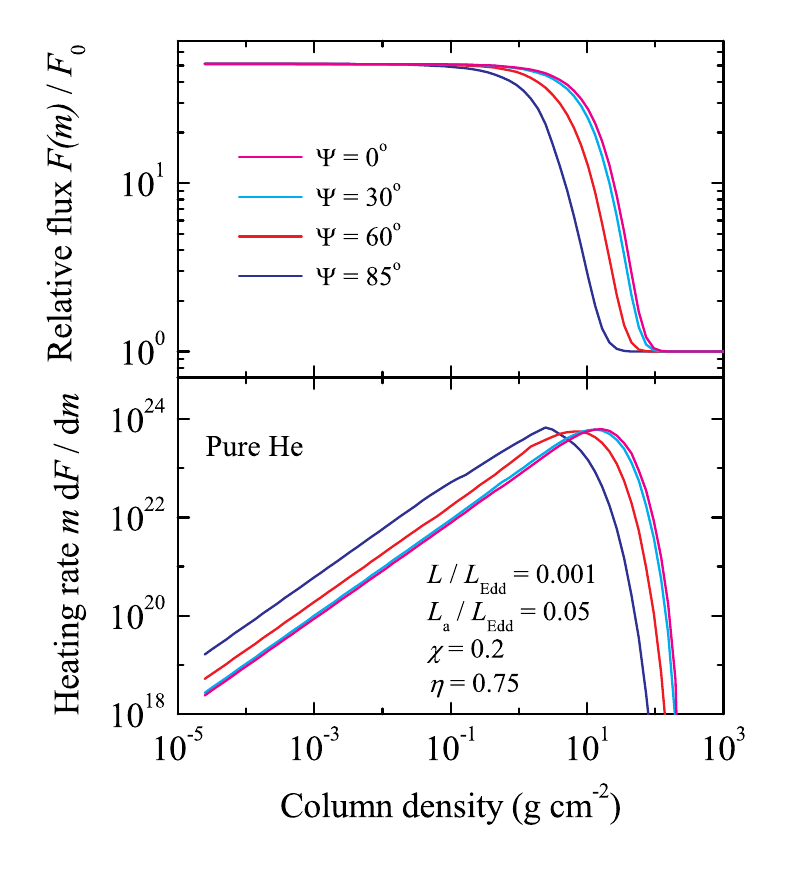}
\caption{\label{fig3}
The bolometric flux (top panel) and the heating rate (bottom panel) as functions of the column density for the models
presented in Fig.\,\ref{fig2}.
}
\end{figure}

 Both forms of the energy balance equation are not satisfied after the first and many following iterations.
The imbalance values for both forms at every column density are used for computation of the correction to the temperature
at every depth \citep[see details in][]{SPW12}. 
All the computations are repeated for the new temperature distribution, and such iterations are repeated to satisfy the energy
 balance at every atmospheric point with the relative flux accuracy  better than 1\%. 

\subsection{Comparison with previous works}

The main difference between the model presented above and the models used by other authors is the inner boundary condition.
We considered atmospheres with a given intrinsic radiative flux $F_0$, while \citet{DDS01} and \citet{Turolla.etal:94} assumed that it is absent, $F_0=0$. 
Formally, \citet{Zampieri.etal:95} introduced the intrinsic luminosity as a parameter, but in fact, they put the inner boundary slightly above the penetration depth of fast particles and did not give the value for the intrinsic luminosity. 
We believe it means that the intrinsic luminosity is small in comparison with the accretion luminosity.  
Therefore, we can compare the previously computed models only with our models at low intrinsic luminosity.

\section{Results}

\label{sec:results}

The physical approach presented in Sect.~\ref{sec:method} was incorporated into the existing stellar atmosphere code described in \citet{SPW12}. 
Using this code we computed a set of models for non-magnetized NS atmospheres heated by accretion. 
The NS mass and radius were fixed at $M$=1.663\,$M_\odot$ and $R$=12\,km, corresponding to the gravitational redshift $1+z = 1.27$, the  surface gravity $\log g = 14.3$, the free fall velocity $V_{\rm ff} = (2GM/R)^{1/2}=1.92\times 10^{10}$\,cm\,s$^{-1} = 0.64\,c$ and the virial temperature $T_{\rm vir} \approx \bar A\,\times7.48\times10^{11}$\,K $\approx \bar A\, 67$\, MeV at the NS surface. 
Here $\bar A$  is the average ion mass in units of the proton mass; $\bar A = 1$ for pure hydrogen matter, $\bar A = 4$ for  pure helium matter, and $\bar A \approx 1.3$ for a solar H/He mix. 
The Eddington luminosity also depends on the chemical composition and can be determined once the hydrogen mass fraction $X$ is fixed, $L_{\rm Edd} \approx 3.8\times10^{38}\,(1+X)^{-1}$\,erg\,s$^{-1}$; $X= 0.7374$ for the solar H/He mixture.
The varied parameters are the relative internal luminosity $\ell = L/L_{\rm Edd}$, the relative accretion luminosity $\ell_{\rm a} = L_{\rm a}/L_{\rm Edd}$, the angle between the bulk velocity vector of the accreted ions and the surface normal $\Psi$,  the bulk velocity of the accreted ions with respect to the free fall velocity $\eta= V_0/V_{\rm ff}$, and the relative temperature of the accreted ions $\chi = T_{\rm i}/T_{\rm vir}$.

\subsection{Atmospheres of low intrinsic luminosity}

Let us first consider a model atmosphere with the low intrinsic luminosity, $\ell=0.001$, and with a relatively low accretion luminosity, $\ell_{\rm a} = 0.05$, which is, however, fifty times larger than the intrinsic luminosity. 
For simplicity, we { consider a} pure helium atmosphere heated by accreted $\alpha$-particles.  
As the fiducial set of parameters, we chose  the temperature of the inflowing ions, $\chi=0.2$, and their bulk velocity $\eta=0.75$.
As the first example, we consider models with { different}  angles of the incoming flow  $\Psi$= 0\degr, 30\degr, 60\degr, and 85\degr. 
The emergent spectra and the temperature structures for these four models are shown in Fig.\,\ref{fig2}. 
Almost all the energy is released at the column density of a few g\,cm$^{-2}$ (Fig.\,\ref{fig3}, bottom panel) and the total radiative flux grows rapidly at those depths (Fig.\,\ref{fig3}, top panel). 
The column density of the maximum energy release depends on the angle $\Psi$:  the smaller the angle the deeper the energy release maximum. 
The atmosphere heated by the fast particles can be divided into an optically thick { inner part} ($m \geq 1 -10$\,g\,cm$^2$) and an optically thin outer part. 
The plasma temperature in the optically thick, almost isothermal part is slightly higher in comparison with that of the undisturbed atmosphere and the radiation flux is controlled by the temperature gradient.  
We can see that this part of the atmosphere is hotter for smaller angles $\Psi$.
 At small column densities, the atmosphere becomes optically thin and the energy generation here cannot be balanced with thermal radiation by plasma with $kT \approx kT_{\rm eff}$ and these layers are overheated. 
 As a result their energy balance is determined  by Compton scattering.

 \begin{figure}
\centering
\includegraphics[angle=0,scale=1.0]{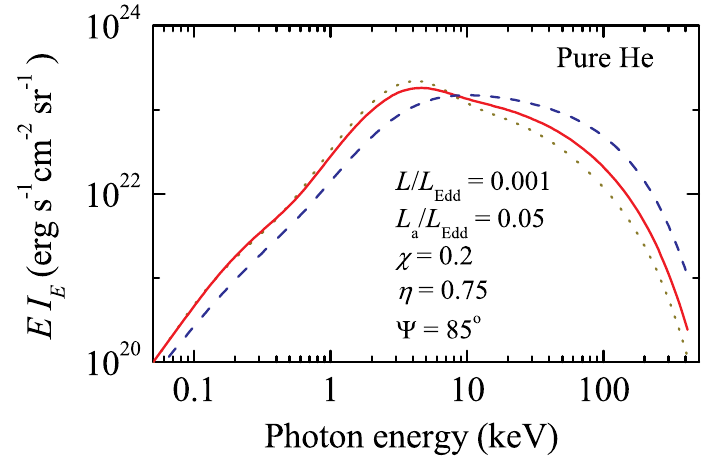}
\caption{\label{fig4}
The emergent specific intensity spectra of low-luminosity ($\ell=0.001$)
helium atmosphere model heated by $\alpha$-particles with accretion luminosity $\ell_{\rm a}=0.05$ 
at three zenith angles of 27\fdg5 (dotted curve), 60\degr (solid curve), and 83\fdg5 (dashed curve).
The  parameters of the model are $\Psi=85\degr$, $\eta=0.75$ and $\chi=0.2$. 
}
\end{figure}

\begin{figure}
\centering
\includegraphics[angle=0,scale=1.0]{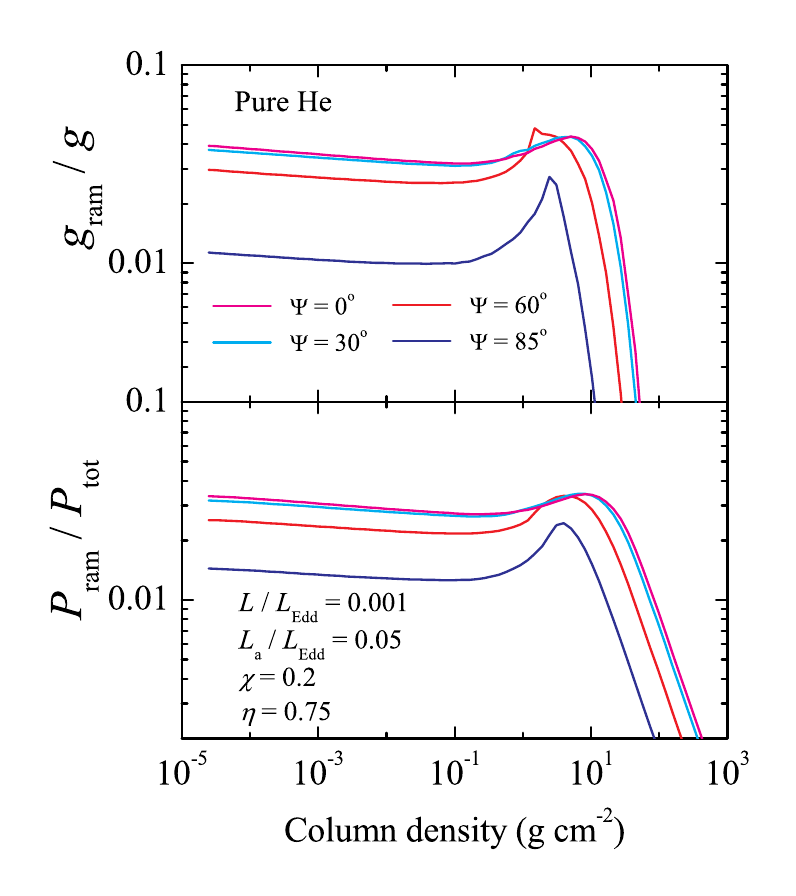}
\caption{\label{fig5}
The relative ram pressure acceleration (top panel) and the relative ram pressure (bottom panel) as functions of the column density for the models presented in Fig.\,\ref{fig2}.
}
\end{figure}

The heating rate of the upper atmospheric layers  by the accreted particles increases with increasing angle $\Psi$. 
Therefore,  models with the  larger  angle $\Psi$ have higher temperatures of the upper layers and they transit to overheated states at  larger $m$. 
The upper layers of the atmosphere with  $\Psi=85\degr$  are heated up to 10$^9$\,K, and the emergent spectrum is close to the Comptonized spectrum of a hot electron slab with Thomson { optical} depth of about unity. 
The emergent flux can be fitted by a power law with exponential cutoff, $F_E \propto E^{ -(\Gamma-1)}\exp(-E/E_{\rm cut})$, at photon energies larger than 5 keV (Fig.\,\ref{fig2}, top panel). 
The photon energy cutoff $E_{\rm cut}= 85$\,keV { is} approximately equal to the surface temperatures $kT$.  
On the other hand, the  blackbody component with a  temperature close to the temperature of the heated optically thick part of the atmosphere dominates the spectra of the models with small $\Psi$, and the optically thin layers add some high energy tails  to the spectra. 
The spectrum of the model with $\Psi=60\degr$ has some intermediate shape with a clear blackbody component and a significant hard tail. 
 Such division of the emergent spectra into two components is obvious in the low-luminosity models computed by  \citet{AW73} and \citet{DDS01}. 
The temperature structure of these models as well as of those computed by \citet{Zampieri.etal:95} show an almost isothermal inner heated slab and the overheated Compton-cooled upper layers with surface temperatures between 10$^8$ and 10$^9$\,K. 

We note also, that the emergent spectra have significant excess at low photon energies in comparison with the blackbody. 
The atmospheres are very opaque at low energies due to the free-free opacity and the emergent radiation at these energies therefore forms  in the overheated upper layers with the escaping monochromatic fluxes close to the blackbody flux with temperatures of those layer.  
This effect was mentioned by \citet{DDS01} as the ``inverse photosphere effect''.      
 
The angular distribution of the emergent spectra is also very unusual for stellar spectra (see Fig.\,\ref{fig4}). 
The normal limb darkening  at the energies below 10 keV alternates with the limb brightening at higher energies,
 as it was suggested earlier by \citet{PG03} for the accreting millisecond pulsars. 
If the bulk velocity vector of the inflowing particles is highly inclined  to the atmosphere normal, their contribution 
to the ram pressure gradient and to the ram pressure itself is small in comparison with that for the particle inflowing along the normal (see Fig.\,\ref{fig5}).

\begin{figure}
\centering
\includegraphics[angle=0,scale=1.0]{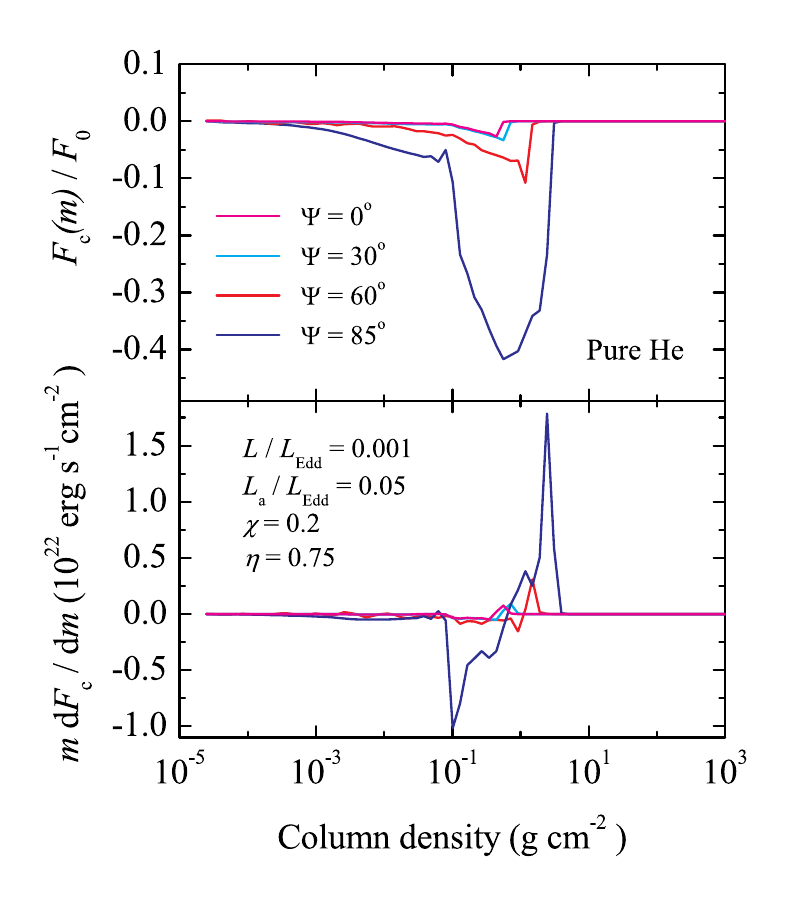}
\caption{\label{fig6}
The relative heat conduction flux (top panel) and the heat conduction flux derivative (bottom panel) as functions of 
the column density for the models presented in Fig.\,\ref{fig2}.
}
\end{figure}

We took into account the thermal conductivity flux in the models, but it is not significant. 
The relative heat flux does not exceed a few tenth of percent in comparison with the  intrinsic radiative  flux even for the atmosphere with the hottest upper layers  (see Fig.\,\ref{fig6}, top panel). 
The contribution of  the heat conduction flux derivative into the differential form of the energy conservation law 
(see Eqs.\,\ref{eq:econs}, \ref{Qc}, \ref{allQ})  is also insignificant  (see bottom panel of Fig.\,\ref{fig6}).
However, the energy generation rate by accretion in the upper layers is small and comparable with the corresponding rate
 provided by conduction  (compare bottom panel of Fig.\,\ref{fig6} to that of Fig.\,\ref{fig3}).
In spite of the fact that the atmosphere is nearly isothermal, the heat flux by conduction becomes here significant, because of the strong temperature dependence (see Eqs. (\ref{eq:fluxc}) and (\ref{eq:kappac})). 
In this case, small numerical fluctuations in temperature of random signs lead to large derivatives of the conduction flux, comparable to the accretion heating. 
This leads to  numerical and, possibly, physical instability of the upper layers where $m < 10^{-4}$\,g\,cm$^{-2}$. 
We suppressed this instability in our computations, but it has to be studied in future investigations.

\begin{figure}
\centering
\includegraphics[angle=0,scale=1.0]{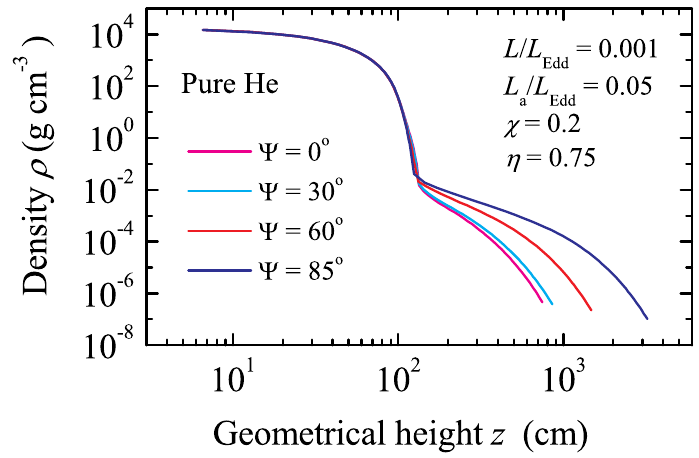}
\caption{\label{fig7}
The atmosphere density dependence on the geometrical height for the models presented in Fig.\,\ref{fig2}.
}
\end{figure}

\begin{figure}
\centering
\includegraphics[angle=0,scale=1.0]{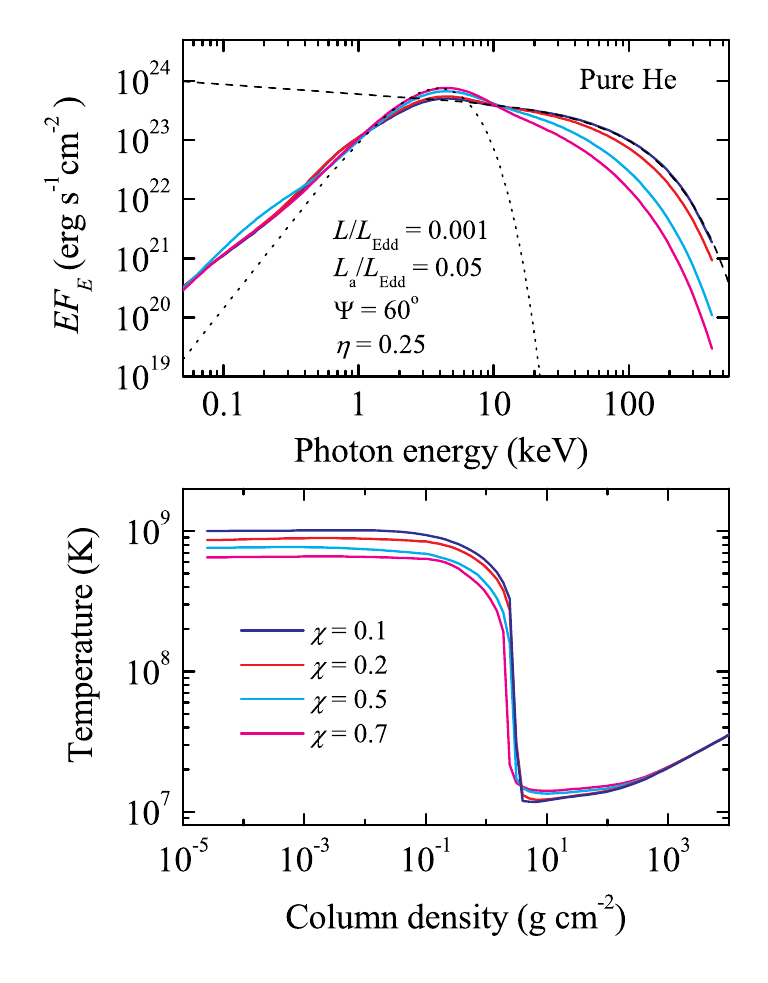}
\caption{\label{fig8}
The emergent energy spectra (top panel) and the temperature structures (bottom panel) of four low luminosity ($\ell=0.001$)
helium model atmospheres heated by $\alpha$-particles with luminosity $\ell_{\rm a}=0.05$ and
parameters $\eta$=0.25 and $\Psi=60\degr$. 
The presented models have different temperature parameter $\chi=$0.1, 0.2, 0.5, and 0.7. 
The fit  to the spectrum above 5 keV with an exponential cutoff power law with  $\Gamma=2.15$, $E_{\rm cut}=85$ keV is shown for   $\chi=$0.1.  
The dotted curve shows the blackbody with temperature $kT=1$ keV.   
}
\end{figure}

Despite the high temperatures of the upper layers, the computed atmospheres are geometrically thin,  as it was mentioned by other authors before 
\citep[see, e.g.][]{Turolla.etal:94},  with a thickness of a few tens of meters at most  (see Fig.\,\ref{fig7}). 
It is not surprising because  the highest model temperature ($\approx 10^9$\,K) is significantly lower than the virial temperature for the protons. 
However, the physical picture can potentially change if the electron-positron pair production is taken into account  \citep{Zane.etal:98}. 
The virial temperature for the electron-positron pairs is close to $10^9$\,K, therefore, we can expect an outflow of pairs at some model parameters. 
The effects of pairs on the atmosphere models will be considered in a separate paper.  

\begin{figure}
\centering
\includegraphics[angle=0,scale=1.0]{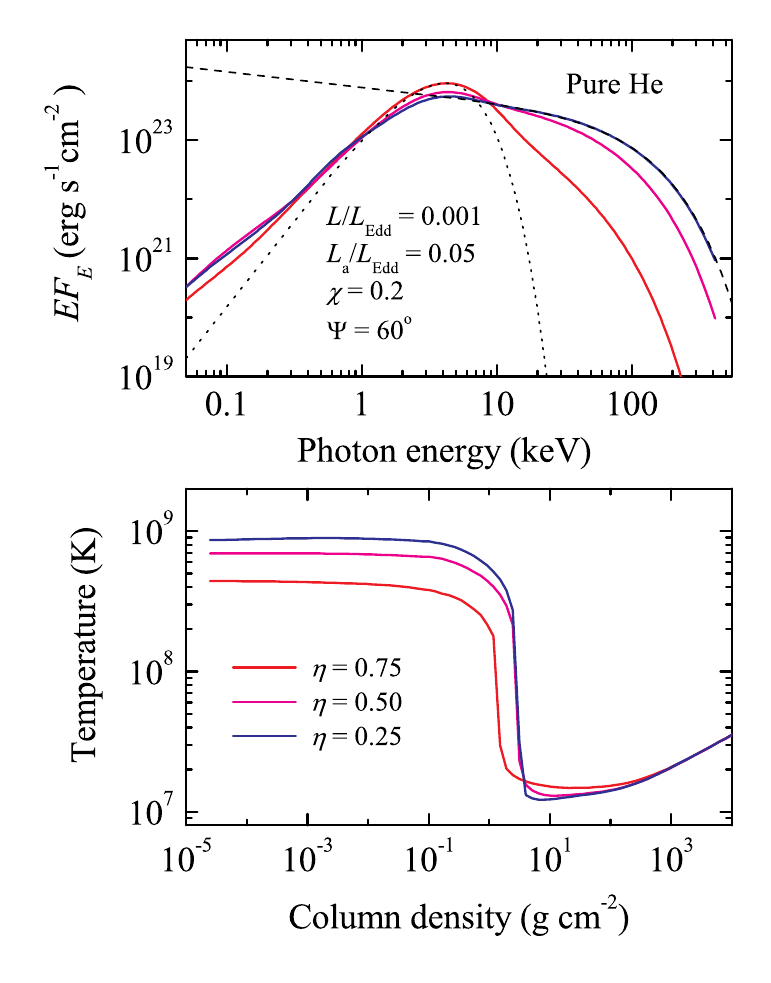}
\caption{\label{fig9}
The emergent spectrum (top panel) and the temperature structure (bottom panel) of the fiducial model  (with $\eta=0.75$) in comparison 
with those of the models with smaller bulk velocities of the accreting particles  ($\eta = 0.50$ and 0.25). 
The fit to the spectrum above 5 keV with an exponential cutoff power law with  $\Gamma=2.25$, $E_{\rm cut}=80$ keV is shown for  $\eta$ = 0.25. 
The dotted curve shows the blackbody with temperature $kT=1.1$ keV.    
}
\end{figure}
 
\begin{figure}
\centering
\includegraphics[angle=0,scale=1.0]{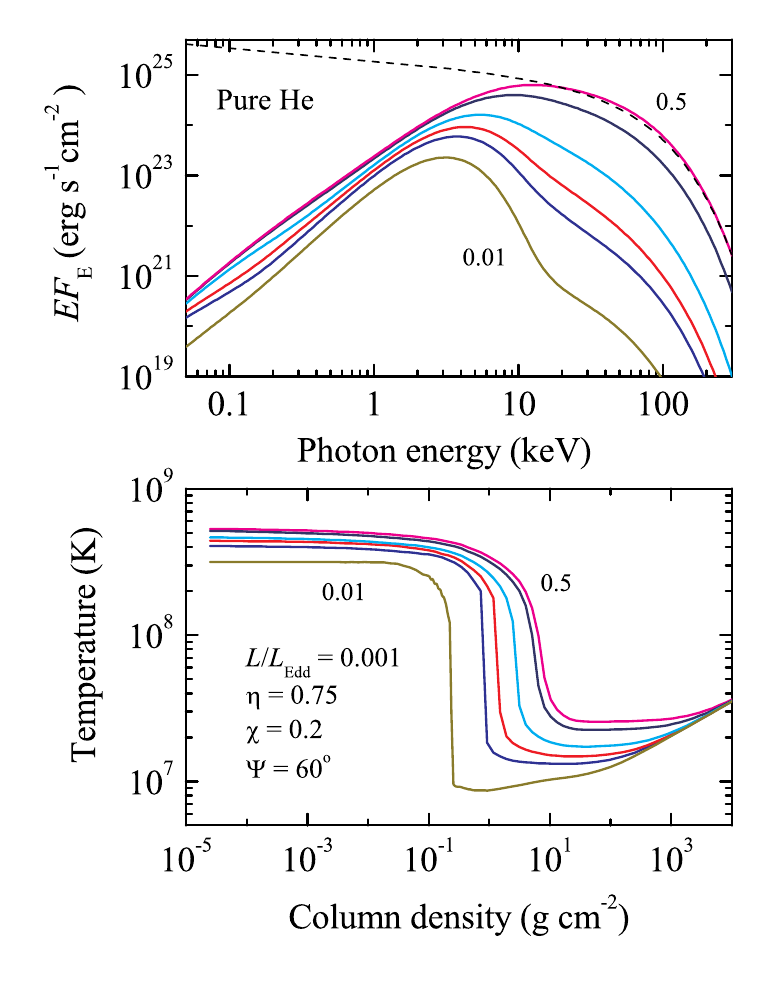}
\caption{\label{fig10}
The emergent spectra (top panel) and the temperature structures (bottom panel) of the models with different  accretion luminosities ($\ell_{\rm a} = 0.01$,  0.03, 0.05, 0.1, 0.3, and 0.5). 
The fit to the spectrum above 15 keV with an exponential cutoff power law with  $\Gamma=2.25$, $E_{\rm cut}=42$ keV is shown for  $\ell_{\rm a}$ = 0.5. 
}
\end{figure}

\subsection{Accretion heated atmospheres with various input parameters}

We have considered above how the physical properties  of the heated atmospheres depend on the incoming angle of the accretion flow.
We now concentrate on the influence of other parameters on the emergent spectrum and the temperature structure. 
We consider pure helium models with two intrinsic luminosities $\ell=0.001$ and  $0.5$, having accretion luminosity $\ell_{\rm a}=0.05$, and other parameters  $\Psi=60\degr$, $\eta=0.75$ and $\chi=0.2$ as in the fiducial model. 
We study then how variations in every input parameter affect the emergent spectrum and the temperature structure. 
 
The obtained results for the intrinsic luminosity $\ell=0.001$ are shown in Figs.\,\ref{fig8}--\ref{fig10}. 
 Varying the temperature and the bulk velocity affects the temperature structures and the emergent spectra of the models 
 in the qualitatively similar way as the inclination angle of the bulk velocity vector made (Figs.\,\ref{fig8} and \ref{fig9}). 
 Such a similarity arises because fast particles penetrate into  different atmospheric depths when we vary their temperature or the bulk velocity. 
 And, therefore, the ratio between the energy, which is released in the optically  thick, heated slab and in the surface overheated
  layer also changes. Indeed, for higher temperature, the fraction of  the particles with high individual velocities increases as well. 
They penetrate into  deeper layers, and the relative contribution of  the optically thick heated layer also increases. 
It becomes hotter, whereas the temperature of the upper overheated layers decreases.
The relative contributions of the blackbody and the Comptonized components also change accordingly (Fig.\,\ref{fig8}). 
In a similar way the particles with the larger bulk velocities  penetrate deeper (Fig.\,\ref{fig9}) and heat the optically thick part of the atmosphere to higher temperatures. 
On the other hand, the accretion heating rate of the very surface layers is larger for the smaller bulk velocities.  

 We note that the dependence of the accretion heated model atmospheres on  the temperature of the fast particles was investigated before by \citet{DDS01}. 
They showed that fast protons of lower temperature heat the overheated upper layers to the higher temperatures, whereas the temperature of the isothermal inner slab is lower than for the corresponding case of high-temperature protons (see their Fig.\,3).
As a result the emergent spectra of the models heated by low-temperature protons are harder. 
Our models (see Fig.\,\ref{fig8}) confirm these results.

The models with varying mass accretion rate have identical dependence of the heating rate over the depth, with the normalizations directly proportional to $\dot m$.  
The  depth of the transition to the overheated parts of the atmosphere and its temperature are determined by the accretion luminosity $\ell_{\rm a}$.  
Higher accretion luminosity leads to higher temperatures of the heated layers, both of the overheated upper layers and of the isothermal region, and to a smoother transition from the isothermal region to the overheated layers. 
The strength of the hard Comptonized  component also depends on $\ell_{\rm a}$. 
At  small accretion luminosities, the blackbody component dominates and the hard tail in the emergent spectra  is relatively weak. 
At high accretion luminosities,  $L_{\rm a} \gg L$, the emergent spectra are dominated by the hard Comptonized component. 
 These results confirm those obtained before by \citet{AW73} and \citet{DDS01}. 
They demonstrated that a rise of the accretion luminosity leads to a harder emergent spectrum and a higher temperature of the overheated layers and the isothermal slabs. 
This fact is especially clear from Fig.\,4 in \citet{DDS01}.

\begin{figure}
\centering
\includegraphics[angle=0,scale=0.95]{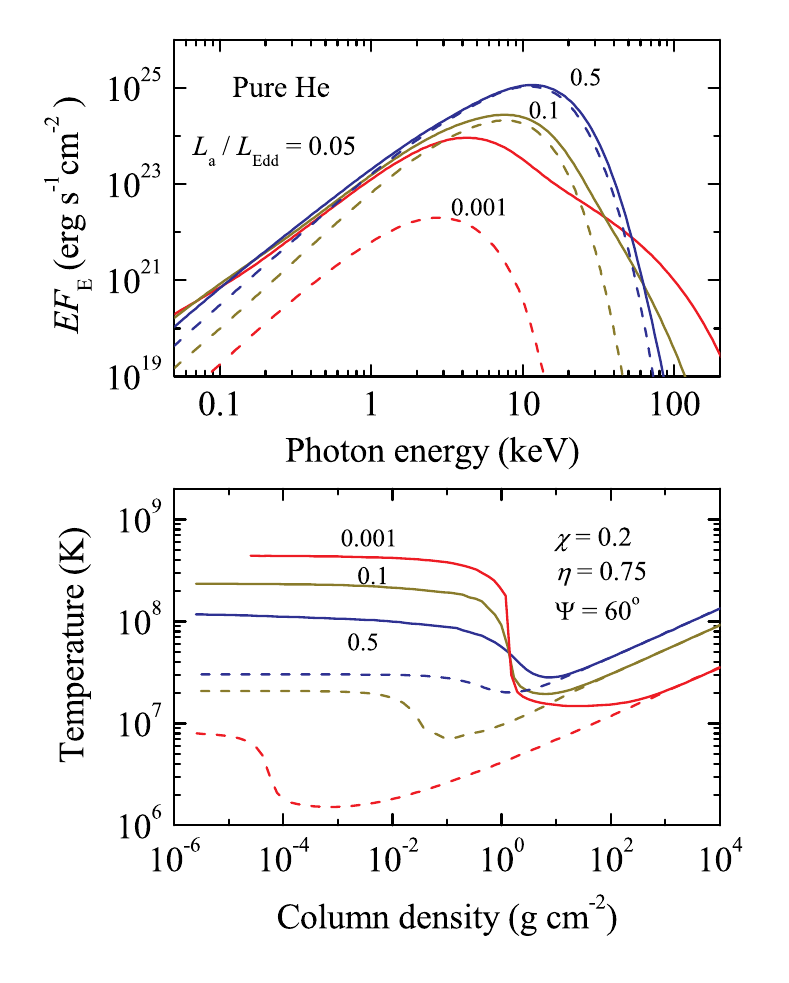}
\caption{\label{fig11}
The emergent spectra (top panel) and the temperature structures (bottom panel) of the models with different intrinsic luminosities
($\ell = 0.5$, 0.1,  and 0.001), but the same accretion luminosity  $\ell_{\rm a}=0.05$. 
The corresponding models for the undisturbed atmosphere  are shown by the dashed curves. 
}
\end{figure}

The influence of the accretion heating decreases when the intrinsic stellar luminosity increases (Fig.\,\ref{fig11}).
The spectrum of the model with the lowest intrinsic luminosity $\ell=0.001$ is completely determined by the heated part of the atmosphere whereas the spectrum of the model with the highest intrinsic luminosity $\ell=0.5$ differs from the spectrum of the undisturbed model very little.

\begin{figure}
\centering
\includegraphics[angle=0,scale=0.95]{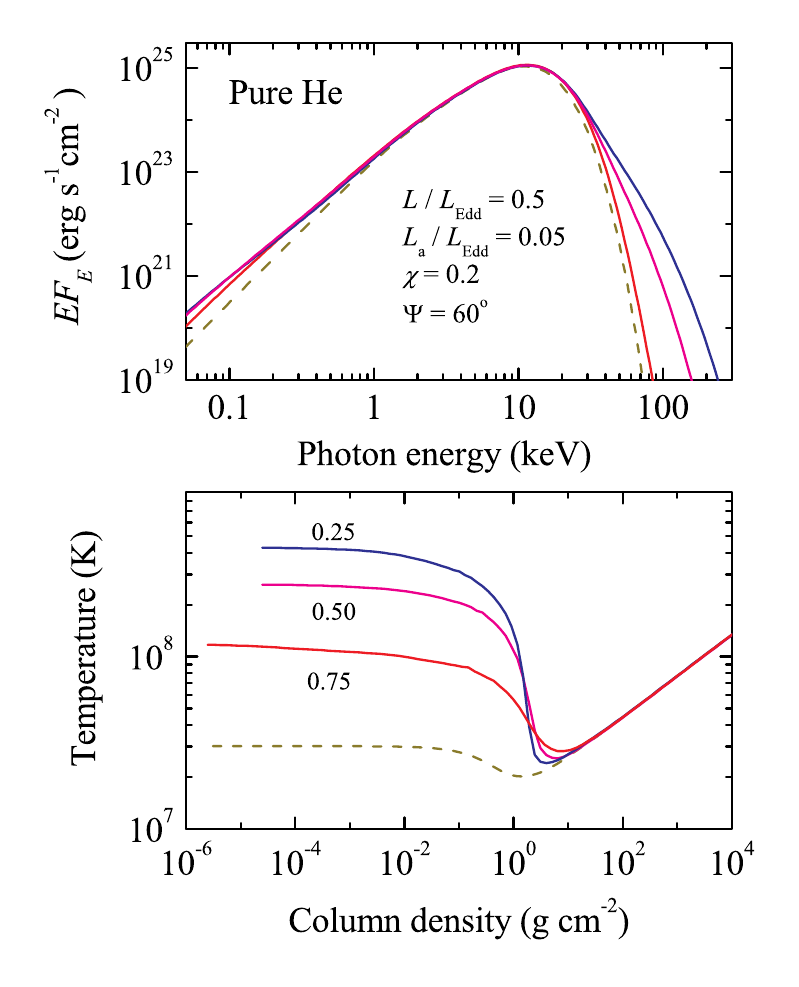}
\caption{\label{fig12}
Same as Fig.~\ref{fig9} but for higher intrinsic luminosity $\ell = 0.5$. 
At the bottom panel, the curves are labeled with the corresponding values of parameter $\eta$.
The spectrum and the temperature structure of the undisturbed model are also shown by the dashed curves. 
}
\end{figure}

We also investigated the influence of the parameters for the case with the high intrinsic luminosity $\ell=0.5$.
The qualitative dependences for the temperature structures are the same as they were described for the models with the low intrinsic luminosity $\ell=0.001$. 
However, the spectra are less sensitive to the variations of the parameters (see Fig.\,\ref{fig12}).

\subsection{Atmospheres with the solar composition}

\begin{figure}
\centering
\includegraphics[angle=0,scale=0.95]{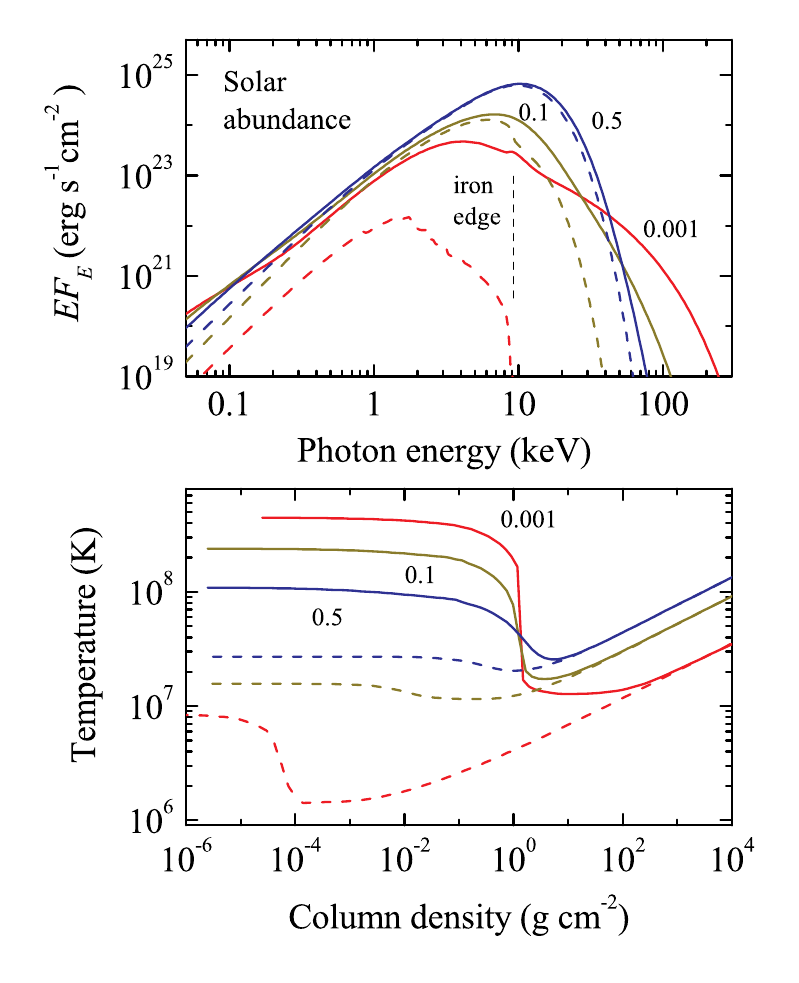}
\caption{\label{fig13}
The emergent spectra (top panel) and the temperature structures (bottom panel) of the models with  solar chemical abundance
and three intrinsic luminosities, $\ell = 0.5$, 0.1 and  0.001, but with the same parameters of the accreting particles
($\ell_{\rm a} = 0.05$, $\eta$=0.75, $\chi=0.2$, $\Psi$=60\degr). 
The spectrum and the temperature structure of the undisturbed models are also shown by the dashed curves. 
}
\end{figure}

The accreting matter in LMXBs has helium rich composition  only in the ultracompact systems  (for instance, in 4U\,1820$-$30). 
Most of the systems have solar chemical composition of the accreting matter.  
Therefore, we have to consider how a change in the chemical composition affects the results. 
As can be seen from Eq.\,(\ref{dvdz}),  protons and $\alpha$-particles decelerate the same way \citep{BSW92}, because the velocity derivative depends on the combination $Z^2/A$ only. 
Here we consider only these two types of fast particles and ignore the nuclei of heavy elements. 
The solar abundance model atmospheres heated by the solar mix of the protons and $\alpha$-particles are similar to the pure helium models (Fig.\,\ref{fig13}).

\begin{figure}
\centering
\includegraphics[angle=0,scale=1.0]{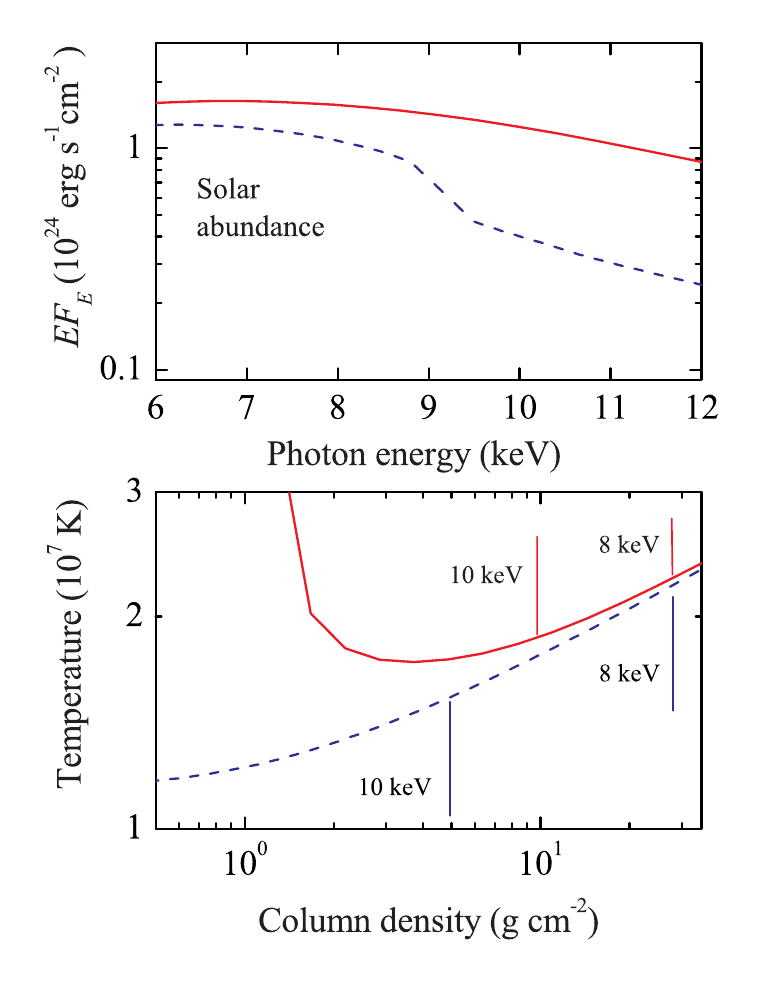}
\caption{\label{fig14}
Magnification of  Fig.\,\ref{fig13}  but for one intrinsic luminosity $\ell = 0.1$ only. 
The { formation} depths of the photons with energies of 8 and 10 keV { are indicated} by thin vertical lines for the heated and the undisturbed atmospheres.
}
\end{figure}

However, there is  some qualitative difference in the emergent spectra.  
The spectra of the low-luminosity undisturbed model atmospheres show absorption edges. 
The photoionisation edge of hydrogen-like iron is the most prominent among them (see model spectra with  $\ell=0.001$ and 0.1 in Fig.\,\ref{fig13}).
The opacity at  photon energies above the photoionisation threshold is larger than that below the threshold. 
Therefore photons above the threshold escape from smaller column densities, where the temperature is lower. 
This is the reason why absorption edges exist. 
In the heated atmospheres, iron is more ionized and the jump in the opacity across the photoionisation threshold is smaller. 
Furthermore, the atmosphere is nearly isothermal  in the region where the photons around the edge are formed (see bottom panel of Fig.\,\ref{fig14}).  
As a result, the spectra of the accretion-heated atmospheres are almost featureless (see top panel of Fig.\,\ref{fig14}).  

\begin{figure}
\centering
\includegraphics[angle=0,scale=1.]{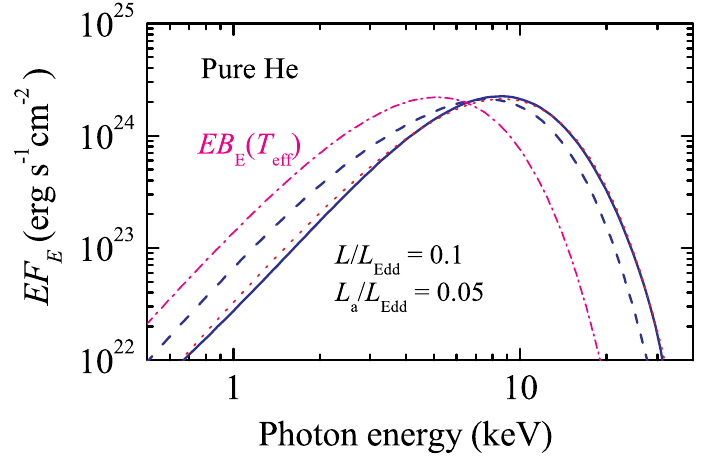}
\includegraphics[angle=0,scale=1.]{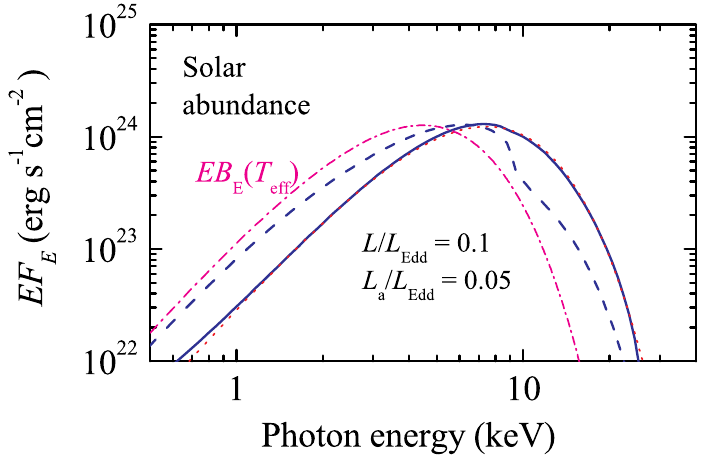}
\caption{\label{fig15}
The residual spectra (solid curves) of the heated model atmospheres with parameters $\ell=0.1, \ell_{\rm a}=0.05$, 
$\chi=0.2$, $\eta=0.75$, and $\Psi=60\degr$ for two chemical compositions, pure helium (top panel) and solar abundance (bottom panel). 
The spectra of undisturbed atmospheres are shown by dashed curves.
The best-fit diluted blackbody spectra to the residual spectra are shown by the dotted curves, 
and the blackbodies of the effective temperatures are shown by the pink dash-dotted curves. 
}
\end{figure}

 \subsection{Application to X-ray bursting neutron stars}

Often the persistent spectra of X-ray bursting NSs are subtracted from the spectra detected during X-ray bursts for obtaining ``pure'' burst spectra   \citep[see, however, ][]{Worpel13,Worpel15,Degenaar16,Kajava17_4U1728}.  
These spectra are well fitted \citep{Galloway08} with  a blackbody  
\be \label{fit_obs}
     f_{E'} \approx  K \pi B_{E'}(T_{\rm BB}),
\ee
where the normalization $K$ and the color temperature $T_{\rm BB}$ are the fitting parameters. 
On the other hand, the observed flux is determined by the model atmosphere flux $F_{E}(T_{\rm eff})$
\be
        f_{E'} = \frac{F_{E}(T_{\rm eff})}{(1+z)^3}\frac{R^2(1+z)^2}{D^2}, 
\ee
where $D$ is the distance to the source, and $z$ is the gravitational redshift,
\be \label{eq:redshift_def}
    1+z=(1-R_{\rm S}/R)^{-1/2}.
\ee
Here $R_{\rm S} = 2GM/c^2$ is the Schwarzschild radius. We note also that $E=E'(1+z)$.
The model spectra of hot NS atmospheres  are close to diluted blackbodies because of the effective interaction between plasma and radiation by Compton scattering  \citep{London86,Lapidusetal:86}.
They can be fitted as
\be \label{fit_mod}
       F_{E} \approx w \pi B_{E} (\fc T_{\rm eff})
\ee
with two fitting parameters,  the color correction factor $\fc$ and the dilution factor $w\approx \fc^{-4}$.
Both fitting parameters depend on the relative luminosity of the NS $\ell=L/L_{\rm Edd}$. 
Therefore, the observed spectrum will be defined as follows
\be
      f_{E'} \approx w \pi \frac{B_{E}(\fc T_{\rm eff})}{(1+z)^3}\frac{R^2(1+z)^2}{D^2}
 \ee
 or
\be      \label{fit_tot}
     f_{E'} \approx w\pi B_{E'}\left[\fc T_{\rm eff}(1+z)^{-1}\right]\ \frac{R^2(1+z)^2}{D^2}.
\ee  
Now we can determine the  parameters $K$ and $T_{\rm BB}$  from comparison of Eqs.~(\ref{fit_obs}) and (\ref{fit_tot}):
 \be
     K \approx w \frac{R^2(1+z)^2}{D^2},
      \ee
 and
 \be
      T_{\rm BB} = \fc T_{\rm eff}(1+z)^{-1}.
       \ee
It is clear that  the observed normalization $K$ can change with the observed bolometric flux  at the cooling burst phases only because of the changes in the dilution factor $w$. 
Therefore, $K$ has to change in a similar fashion as the dilution factor $w$ varies with the relative luminosity $\ell$. 
If we account for deviation of the bolometric flux from the observed blackbody flux $F_{\rm BB}$, the observed relation $K-F_{\rm BB}$ should be fitted with the theoretical curve $w - w\fc^4 \ell$. 
Here $(w\fc^4)^{-1}$ is the bolometric correction, as was shown by \citet{Suleimanov.etal:17}.
From such fitting two parameters, the observed Eddington flux $F_{\rm Edd,\infty}$ and the NS observed solid angle $(R(1+z)/D)^2$ can be found, and NS mass and radius can be limited. 
This approach is known as the cooling tail method, see details in \citet{SPW11}, \citet{Poutanen.etal:14}, and \citet{Suleimanov.etal:17}. 
This method would be correct if a passively cooling NS were a correct model for the evolution of  the burst spectra in the cooling phase. 
Indeed, the normalization of the spectra of bursts taking place during the hard persistent states evolve according to this model at  a  relatively high  luminosity $\ell > 0.2-0.7$. 
This allowed us to obtain constraints on the NS masses and radii \citep{SPRW11,Poutanen.etal:14,Nattila.etal:16,Suleimanov.etal:17}.

\begin{figure}
\centering
\includegraphics[angle=0,scale=1.]{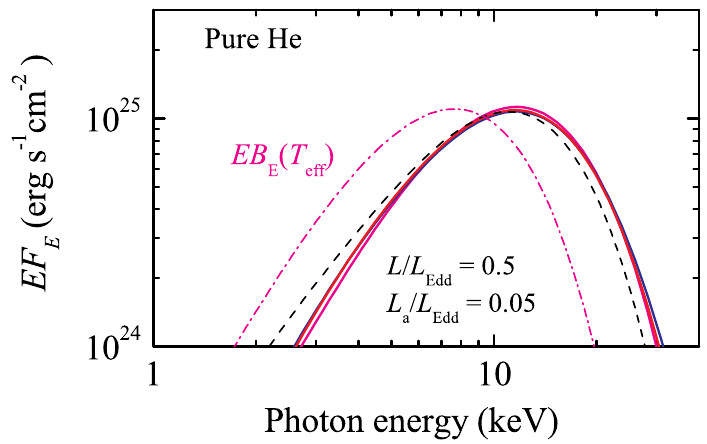}
\caption{\label{fig16}
The residual spectra (solid curves) of the heated helium atmosphere model  with parameters $\ell=0.5, \ell_{\rm a}=0.05$, 
$\chi=0.2$, and $\Psi=60\degr$ for three different values of $\eta=$ 0.75, 0.50, and 0.25 (see Fig.\,\ref{fig12}). 
The spectrum of an undisturbed atmosphere is shown by the blue dashed curve and the blackbody of the effective temperature is shown 
by the pink dash-dotted curve.}
\end{figure}

\begin{figure}
\centering
\includegraphics[angle=0,scale=1.]{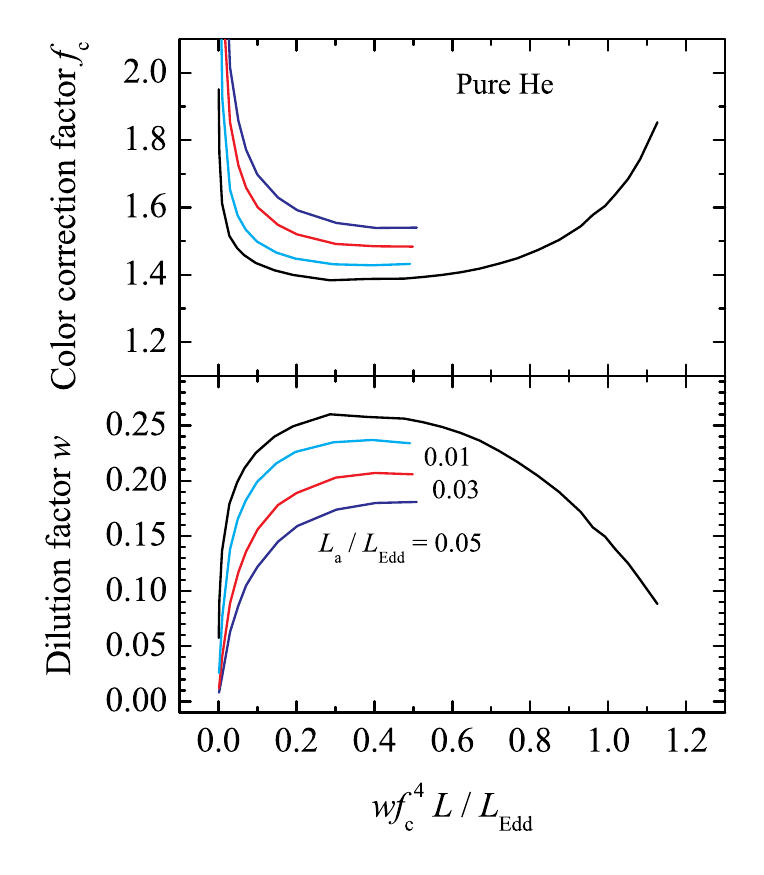}
\caption{\label{fig17}
Color correction (top panel) and dilution (bottom panel) factors as  functions of the corrected relative luminosity. 
They are computed using residual spectra of the heated helium model atmospheres for three accretion rates, $\ell_{\rm a}=$0.05 (dark blue), 0.03 (red), and 0.01 (light blue). 
Other parameters are $\chi=0.2$, $\eta=0.75$, and $\Psi=60\degr$. 
The  corresponding factors computed using spectra of undisturbed atmospheres are also shown by black curves.}
\end{figure}

\begin{figure}
\centering
\includegraphics[angle=0,scale=1.]{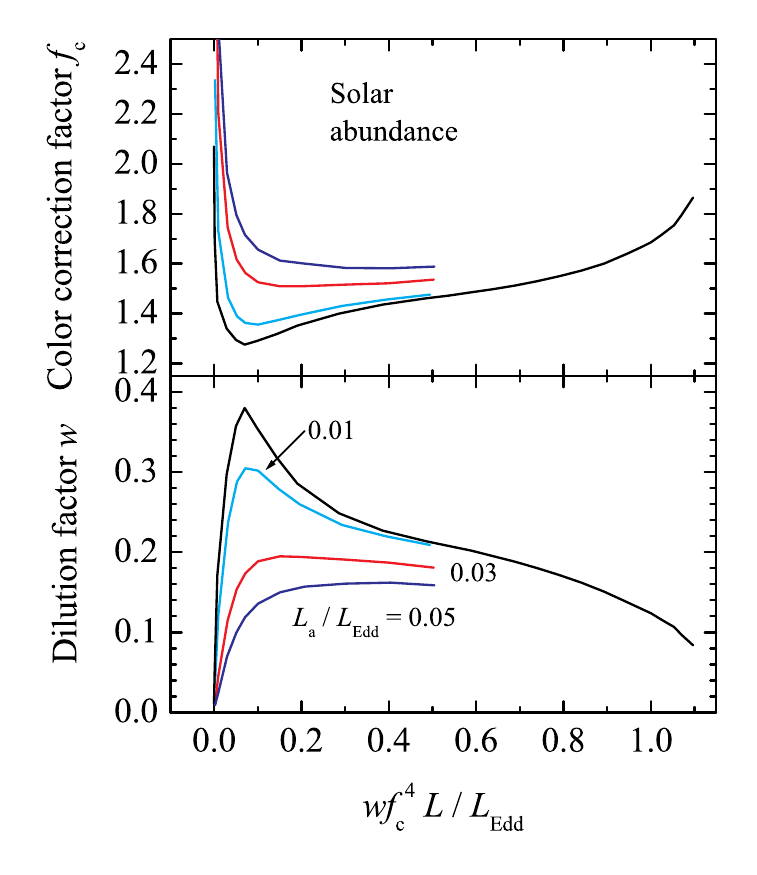}
\caption{\label{fig18}
The same as in Fig.\,\ref{fig17}, but for atmospheres of solar composition. 
Here the accretion flow temperature parameter is $\chi=0.3$.}
\end{figure}

However, the observed dependences $K - F_{\rm BB}$ deviate from the model curves $w - w\fc^4 \ell$ at low $\ell$. 
The amplitude of the deviations depends on the relative persistent luminosity of the system before the X-ray burst. 
Compare, for instance, small deviations of the model curves from the observed ones for the systems with low persistent emission, $L_{\rm per} \approx 0.01 L_{\rm Edd}$ \citep{Nattila.etal:16}, and the large deviations for the system 4U\,1820$-$30 \citep{Suleimanov.etal:17} with a relatively high persistent luminosity, $L_{\rm per} \approx 0.04-0.07 L_{\rm Edd}$.
We suggest that these deviations are determined by additional heating of the NS atmospheres by the accretion flow.
Let us  evaluate the effect of such a heating on the theoretical dependences $w - w\fc^4 \ell$ to study this hypothesis.

To obtain the  model spectra we use the same approach as used for the analysis of the observed spectra of X-ray bursting NSs \citep[see, e.g.][]{Galloway08}. 
We assume that the spectrum of the heated model atmosphere with the lowest intrinsic luminosity $\ell=0.001$ represents  the  spectrum of the persistent emission from an accreting NS. 
We then subtract this spectrum from the spectra corresponding to the higher intrinsic luminosities to mimic pure burst model spectra. 
The examples of such residual spectra are presented in Fig.\,\ref{fig15}. 
The residual spectra can be fitted by diluted blackbodies even better than the spectra of undisturbed model atmospheres and their color temperatures are higher than the corresponding color temperatures of the spectra of undisturbed model atmospheres. 
We also note, that the absorption edge clearly seen in the model spectrum of the solar-abundance undisturbed  atmosphere completely disappears 
in the residual spectrum  (Fig.\,\ref{fig15}, bottom panel). 
In the first approximation the reduced spectra are slightly dependent on the parameters of the  accretion flow (Fig.\,\ref{fig16}).

Using the grid of residual model spectra in the intrinsic luminosity range $\ell$ from 0.001 to 0.5 we computed the color correction and dilution factors  for pure He and solar abundance atmospheres (see Figs.\,\ref{fig17} and \ref{fig18}).  
{ We} do not consider luminosities above $\ell=0.5$, because at those luminosities the radiation pressure should affect significantly the dynamics of the accreted fast particles.   
It is also possible that during X-ray bursts the accretion luminosity  smoothly increases when the burst luminosity decreases. 
We leave the models with high intrinsic  luminosities for future investigations. 

We considered three values of the relative accretion luminosities, $\ell_{\rm a}$=0.05, 0.03 and 0.01. 
We took the accretion flow parameters to be $\eta=0.75$, $\Psi=60\degr$, and $\chi=0.2$ for pure helium models, and $\chi=0.3$ for the solar abundance models. 
The residual spectra were fitted with the diluted blackbody spectra in the blue-shifted observational energy range of the {\it RXTE} observatory, $(3-20)(1+z)$ keV. 
Here the gravitational redshift is $1+z=1.27$ for the accepted NS parameters and $\log g = 14.3$, see Sect.\,\ref{sec:results}. 
According to expectations, the color correction factors are larger for the heated model atmospheres, and their values depend on the mass accretion rate. 
We note that the prominent dip at $\ell \approx 0.1$ in the theoretical $\fc - w\fc^4 \ell$ curve computed for the undisturbed solar-abundance atmospheres becomes weaker for slightly heated model atmospheres ($\ell_{\rm a}=0.01$) and completely disappears in the curves computed for higher accretion luminosities.   

\begin{figure}
\centering
\includegraphics[angle=0,scale=0.7]{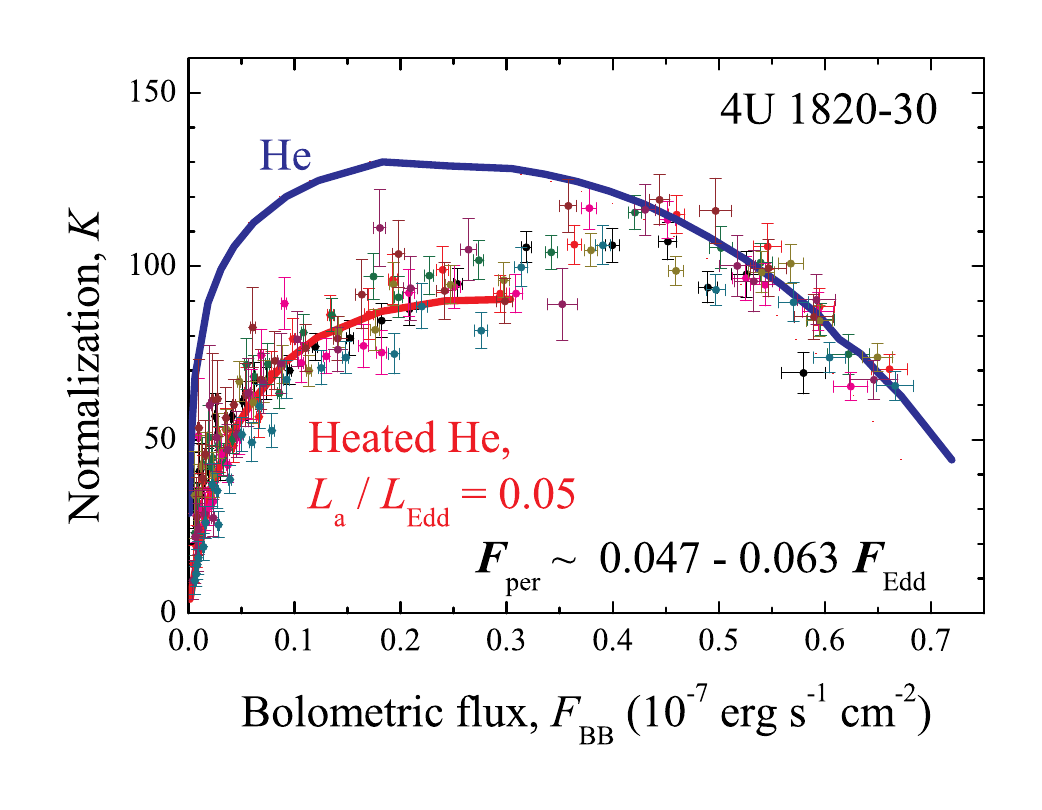}
\includegraphics[angle=0,scale=0.7]{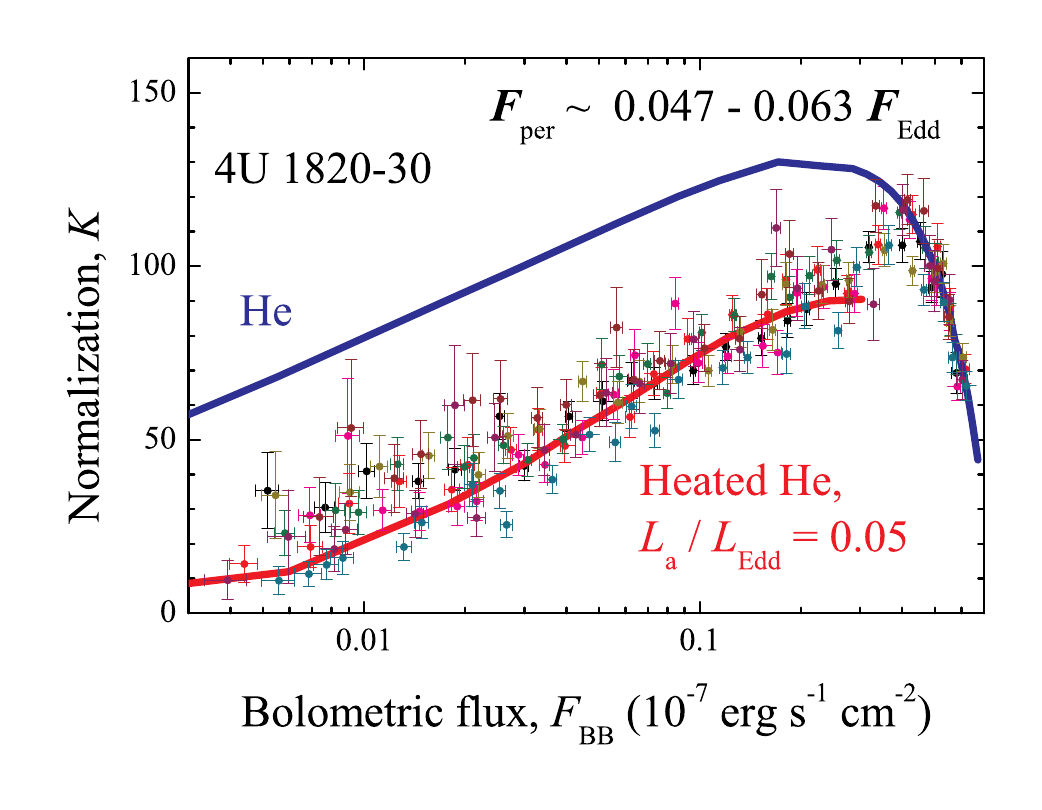}
\caption{\label{fig19}
Comparison of the observed dependences $K - F_{\rm BB}$ obtained for selected X-ray bursts of 4U\,1820$-$30 \citep[see][]{Sul.etal:17} with the model curves $w - w\fc^4 \ell$ computed for undisturbed (blue curves) and heated (red curves)  pure helium atmospheres. 
The axes of the fluxes are shown in linear (top panel) and logarithmic (bottom panel) scales.
The fitting parameters $F_{\rm Edd, \infty}= 0.6\times 10^{-7}$~erg\,s$^{-1}$\,cm$^{-2}$ and $\Omega=500$\,(km/10\,kpc)$^{-2}$ 
are the same for both model curves.
}
\end{figure}

In the previous work \citep{Sul.etal:17} we considered X-ray bursts occurring in a hard state of the ultracompact system 4U\,1820$-$30 which likely accretes helium from a white dwarf companion. 
We applied the direct cooling tail method of \citet{Suleimanov.etal:17} to obtain constraints on the NS mass and radius.
The persistent luminosities  before the considered bursts were high enough (0.047--0.063 $L_{\rm Edd}$) and the observed dependences 
$K - F_{\rm BB}$ deviated from the model curves at $\ell < 0.6$ (see Fig.\,\ref{fig19}). 
Therefore, we analyzed only the high-luminosity part of the observed data. 
We suggested that deviations of the theoretical curve from the data are likely caused by accretion during the later stages of the cooling tail. 
Taking the accretion luminosity of 5\% of the Eddington value ($\ell_{\rm a}=0.05$), the model curves $w - w\fc^4 \ell$  
for the heated pure helium atmospheres shift down sufficiently to pass  through the data points at $\ell < 0.5$.
At the relative luminosities higher than  $\ell \approx$\,0.3, i.e. $F_{\rm BB} \approx$\,0.2$\times 10^{-7}$\, erg\,s$^{-1}$\,cm$^{-2}$, the data lie slightly above the model curve.  
We suggest that when the burst luminosity drops from  $\ell \approx0.6$ to $\approx0.3$, the accretion rate increases causing 
a smooth shift of the blackbody normalization $K$ from a theoretical curve with no accretion $\ell_{\rm a}=0$ to a curve corresponding to  $\ell_{\rm a}=0.05$.  
There are a few possible reasons why the accretion restarts again after some time. 
First, the inner parts of the accretion flow can be destroyed during the photospheric radius expansion phase. 
The typical time of the photosphere to settle down to the NS surface can be much shorter than the viscous time for the inner accretion flow. 
Therefore, the accretion restarts again gradually with some time gap. 
Second, is the influence of radiation pressure.
 The radiation field has two effects on the particle dynamics near the NS \citep{ML93}.  
 At high luminosity, the direct radial radiation force may reduce the radial velocity decreasing mass accretion rate, while at low luminosity 
 the radial velocity may increase as a result of the angular momentum loss due to the Poynting-Robertson drag. 
 Finally, the burst radiation may cool the accretion flow making it thinner. As a result, only a relatively small part of the NS surface is affected by  accretion. 
It may well be possible that all three reasons play a role. 

\section{Summary}

In this paper  we presented a computational method  to construct NS { model} atmospheres heated by accreted particles. 
Our results confirm previous findings  \citep[e.g.][]{AW73, Zampieri.etal:95, DDS01} that  fast heavy particles (protons, $\alpha$-particles, or their mix) penetrate down to some depth in the atmosphere and release most of their  kinetic energy in a relatively narrow layer due to Coulomb interaction with electrons. 
The heating by accretion  is balanced by the cooling due to mainly free-free emission and Comptonization. 
The relatively dense optically thick region, where the cooling is dominated by free-free  emission, is only slightly hotter than the corresponding layer of an undisturbed atmosphere. 
It forms a quasi-isothermal transition region between the inner atmosphere unaffected by accretion heating and the hot ($\sim 10^8 - 10^9$\,K) rarefied upper layer. 
The upper hot layers  cool mainly  by  Compton scattering.  
The width of the transition region and the temperature of the  upper layer depend mainly on the accretion luminosity $L_{\rm a}$ and the intrinsic NS luminosity $L$.

The emergent spectrum  of an accretion-heated NS atmosphere is wider than the spectrum of the corresponding undisturbed atmosphere and can be roughly represented as a sum of two components: a blackbody-like radiation from the transition region and Comptonized (cutoff power-law-like) spectrum of the hot upper layers. 
The relative contribution of the components is determined by the ratio of the intrinsic luminosity $L$ to the accretion luminosity $L_{\rm a}$.  
If $L_{\rm a} \gg L$,  the total spectrum is dominated by the Comptonized spectrum of the upper, hot rarefied layers. 
This kind of spectra are similar to the observed spectra of LMXBs in their low hard spectral states, as mentioned by \citet{DDS01}, and their slope depends on the input parameters of the accreted particles: velocity, temperature,  and the penetration angle. 
We note, however, that all computed spectra have photon indexes $\Gamma > 2$.
The observed spectra of LMXBs in their low hard spectral states often have harder spectra with $\Gamma < 2$. 
It means that the contribution of the accretion flow emission to the total observed spectra may not be negligible.
In the opposite case, when $L_{\rm a} \ll L$,  the total spectra are very similar to the spectra of the undisturbed models with harder Wien tails and increased flux at low photon energies. 
In that case, the final spectra depend very little on the properties of accreted particles.
 
The atmospheres heated by material of solar abundance have an important qualitative feature in comparison to pure helium atmospheres heated by  $\alpha$-particles only.  
The spectra of the relatively low-luminosity ($L \ll L_{\rm Edd}$)  undisturbed model atmospheres show significant absorption edges due to photoionisation of hydrogen-like iron. 
These edges disappear in the spectra of the accretion-heated atmospheres because of a significant change in the  temperature structure.  

We also investigated the influence of the accretion heating on the model curves $w - w\fc^4 \ell$ and $\fc - w\fc^4 \ell$, which are used in the (direct) cooling tail method. 
As it was expected the color-correction factors $\fc$ are larger for the heated atmospheres in comparison with the undisturbed atmospheres, and the dilution factors $w$ are smaller. 
The model curve computed for heated helium atmospheres (with $\ell_{\rm a}=0.05$) is well fitted to the low-luminosity part of the observed data $K-F_{\rm BB}$ obtained for X-ray bursts taken place during the hard state of the helium-accreting system 4U\,1820$-$30 \citep[see][]{Sul.etal:17}.

The model curves $\fc - w\fc^4 \ell$ computed for the heated solar-abundance atmospheres do not have a dip at $\ell \approx 0.1$, which is clearly seen in the model curves computed for the undisturbed solar-abundance atmospheres.
This fact may have important implications for interpretation of the X-ray data on bursting NSs that accrete gas of solar abundance, for example, the Clocked Burster GS\,1826$-$24 \citep[see discussion in][]{ZCG12}. 
We plan to apply the method described here for interpretation of the spectral evolution of that and other bursters in a follow-up paper.

\begin{acknowledgements}  
The authors acknowledge A.\,M. Beloborodov for useful discussions and an anonymous referee for the very insightful comments.
This research has been supported by the grant 14.W03.31.0021 of the  Ministry of Education and Science of the Russian Federation.
V.F.S. also thanks Deutsche Forschungsgemeinschaft (DFG) for financial support (grant WE 1312/51-1). 
We thank the German Academic Exchange Service (DAAD, project 57405000) and the Academy of Finland (project 317552)
 for  travel grants. This work benefited from discussions at the BERN18 Workshop supported by the National Science
 Foundation under Grant No. PHY-1430152 (JINA Center for the Evolution of the Elements).
\end{acknowledgements}


\end{document}